\documentclass[preprint,12pt]{elsarticle}
\usepackage{etex}
\usepackage{graphicx}
\usepackage{amssymb}
\usepackage{lineno}

\usepackage{amsfonts}
\usepackage{amsmath}
\usepackage{amsthm}
\usepackage{psfrag,rotating}

\newtheorem{thm}{Theorem}[section]
\newtheorem{lem}{Lemma}[section]
\newtheorem{cor}{Corollary}

\newtheorem{remark}{Remark}[section]

\usepackage{bm,todonotes}

\usepackage{algorithm}
\usepackage{algorithmic}

\usepackage{color,hyperref}					%Uses hyperlinks and colors them blue
\hypersetup{colorlinks,breaklinks,
           linkcolor=blue,urlcolor=blue,
           anchorcolor=blue,citecolor=blue}

\usepackage[caption=false]{subfig}
\usepackage{float}
\usepackage{comment}
\graphicspath{{fig/}}

\usepackage{pstricks-add}
\newbox{\LegendeRDU}
\savebox{\LegendeRDU}{
    \begin{pspicture}(0,0)(0.6,0)
    \psdots[dotstyle=triangle,linecolor=red,dotsize=5pt,fillstyle=solid,fillcolor=red](0.25,0.1)
    \psline[linewidth=0.03,linecolor=red](0,0.1)(0.5,0.1)
    \end{pspicture}}
    
\newbox{\LegendeBC}
\savebox{\LegendeBC}{
    \begin{pspicture}(0,0)(0.6,0)
    \psdots[dotstyle=o,linecolor=blue,dotsize=5pt,fillstyle=solid,fillcolor=blue](0.25,0.1)
    \psline[linewidth=0.03,linecolor=blue](0,0.1)(0.5,0.1)
    \end{pspicture}}
    
\newbox{\LegendeBDT}
\savebox{\LegendeBDT}{
    \begin{pspicture}(0,0)(0.6,0)
    \psdots[dotstyle=triangle,linecolor=blue,dotsize=5pt,dotangle= 180,fillstyle=solid,fillcolor=blue](0.25,0.1)
    \psline[linestyle=dashed,dash=2.5pt 1.5pt,linewidth=0.03,linecolor=blue](0,0.1)(0.5,0.1)
    \end{pspicture}}
    
\newbox{\LegendeRDDD}
\savebox{\LegendeRDDD}{
    \begin{pspicture}(0,0)(0.6,0)
    \psdots[dotstyle=triangle,linecolor=red,dotsize=5pt,dotangle=180,fillstyle=solid,fillcolor=red](0.25,0.1)
    \psline[linestyle=dashed,dash=2.5pt 1.5pt,linewidth=0.03,linecolor=red](0,0.1)(0.5,0.1)
    \end{pspicture}}

\newbox{\LegendeRSDd}
\savebox{\LegendeRSDd}{
    \begin{pspicture}(0,0)(0.6,0)
    \psdots[dotstyle=square,linecolor=red,dotsize=5pt,dotangle=180,fillstyle=solid,fillcolor=red](0.25,0.1)
    \psline[linestyle=dashed,dash=2pt 0.5pt 1pt 0.5pt,linewidth=0.03,linecolor=red](0,0.1)(0.5,0.1)
    \end{pspicture}}
    
\newbox{\LegendeRDD}
\savebox{\LegendeRDD}{
    \begin{pspicture}(0,0)(0.6,0)
    \psdots[dotstyle=diamond,linecolor=red,dotsize=5pt,dotangle=180,fillstyle=solid,fillcolor=red](0.25,0.1)
    \psline[linestyle=dashed,dash=2.5pt 1.5pt,linewidth=0.03,linecolor=red](0,0.1)(0.5,0.1)
    \end{pspicture}}

\journal{}

\begin{document}

\begin{frontmatter}



\title{On Polynomial Chaos Expansion via Gradient-enhanced $\ell_1$-minimization}

\author[label1]{Ji Peng}
\author[label2]{Jerrad Hampton}
\author[label2]{Alireza Doostan\corref{cor1}}
\ead{alireza.doostan@colorado.edu}

\cortext[cor1]{Corresponding Author: Alireza Doostan}

\address[label1]{Mechanical Engineering Department, University of Colorado, Boulder, CO 80309, USA}
\address[label2]{Aerospace Engineering Sciences Department, University of Colorado, Boulder, CO 80309, USA}

%%%%%%%%%%%%%%%%%%%%%%%%%%%%%%%%%%%%%%%%%%%%%%%%%%%%%
%
\begin{abstract}
%
%%%%%%%%%%%%%%%%%%%%%%%%%%%%%%%%%%%%%%%%%%%%%%%%%%%%%
Gradient-enhanced Uncertainty Quantification (UQ) has received recent attention, in which the derivatives of a Quantity of Interest (QoI) with respect to the uncertain parameters are utilized to improve the surrogate approximation. Polynomial chaos expansions (PCEs) are often employed in UQ, and when the QoI can be represented by a sparse PCE, $\ell_1$-minimization can identify the PCE coefficients with a relatively small number of samples. In this work, we investigate a gradient-enhanced $\ell_1$-minimization, where derivative information is computed to accelerate the identification of the PCE coefficients. For this approach, stability and convergence analysis are lacking, and thus we address these here with a probabilistic result. In particular, with an appropriate normalization, we show the inclusion of derivative information will almost-surely lead to improved conditions, e.g. related to the null-space and coherence of the measurement matrix, for a successful solution recovery. Further, we demonstrate our analysis empirically via three numerical examples: a manufactured PCE, an elliptic partial differential equation with random inputs, and a plane Poiseuille flow with random boundaries. These examples all suggest that including derivative information admits solution recovery at reduced computational cost.

\end{abstract}
\begin{keyword}
Polynomial chaos \sep Uncertainty quantification  \sep Stochastic PDEs \sep Compressive sampling \sep $\ell_1$-minimization \sep Gradient-enhanced $\ell_1$-minimization \sep Sparse approximation
%%% keywords here, in the form: keyword \sep keyword
%%% MSC codes here, in the form: \MSC code \sep code
%%% or \MSC[2008] code \sep code (2000 is the default)
\end{keyword}
\end{frontmatter}

\section{Introduction}
\label{sec:intro}
In complex engineering system analysis, inherent variability in inputs and imperfect knowledge of the physics can lead to unfounded confidence or unnecessary diffidence in understanding Quantities of Interest (QoI). Uncertainty Quantification (UQ)~\cite{Ghanem91a,LeMaitre10,Xiu10a} as a study aims to develop numerical tools that can accurately predict QoI and facilitate the quantitative validation of the simulation model.

To characterize uncertainty, probability is a natural framework. We model the uncertain inputs as a $d-$dimensional vector of independent random variables $\bm\Xi:=(\Xi_1,\dots,\Xi_d)$, with probability density function $\rho(\bm\Xi)$. The QoI that we seek to approximate is denoted by a scalar $u(\bm\Xi)$. Here we utilize polynomial chaos expansions (PCEs)~\cite{Ghanem91a,Xiu02} to approximate $u(\bm\Xi)$, assumed to have finite variance. In this case, $u(\bm\Xi)$ can be represented as an expansion in multivariate orthogonal polynomials $\psi_j(\bm\Xi)$, i.e.,
\begin{equation}
u(\bm\Xi)=\sum_{j=1}^{\infty}c_j\psi_j(\bm\Xi)\approx\sum_{j=1}^{P}c_j\psi_j(\bm\Xi)+\epsilon_t(\bm{\Xi}),
\label{eq:PCE}
\end{equation}
where $c_j, j=1,2,\dots$, are the corresponding PCE coefficients, and $\epsilon_t$ is the truncation error associated with retaining $P$ terms of a sorted basis. The PCE coefficients can be computed by the projection
\begin{equation}
c_j = \int u(\bm\Xi)\psi_j(\bm\Xi)\rho(\bm\Xi)d\bm{\Xi}=\mathbb{E}\left[u(\bm{\Xi})\psi_j(\bm{\Xi})\right],
\end{equation}
where the operator $\mathbb{E}$ denotes the mathematical expectation. Here we assume that $\psi_j(\bm\Xi)$ are normalized such that $\mathbb{E}\left[\psi_j^2(\bm{\Xi})\right]=1$.

Typically, a full $P$ term approximation is not necessary, and we can restrict to an unknown subset $\mathcal{C}\subset\{1,\dots,P\}$ such that $|\mathcal{C}|$, the number of elements in $\mathcal{C}$, is significantly smaller than $P$. In addition, $|\mathcal{C}|$ is referred to as the sparsity of the approximation. We approximate $u(\bm\Xi)$ in \eqref{eq:PCE} then by
\begin{equation}
u(\bm\Xi)\approx \sum_{j\in\mathcal{C}}c_j\psi_j(\bm\Xi).
\label{eq:PCA}
\end{equation}
This concept of using one basis, indexed by $\{1,\dots,P\}$, to compute an approximation in a small but unknown subset of those basis functions, indexed by $\mathcal{C}$, falls within the context of compressive sampling~\cite{Candes06a, Donoho06b, Doostan11a, Peng14, Hampton14}.

To identify PCE coefficients, we consider non-intrusive sampling methods, in which deterministic solvers for the QoI are not modified. Such methods include Monte Carlo simulation~\cite{Reagan03,LeMaitre10}, pseudo-spectral stochastic collocation~\cite{Mathelin03,Xiu05a,LeMaitre10, Constantine12a}, least squares regression~\cite{Hosder06,Migliorati13,Hampton15}, and $\ell_1$-minimization~\cite{Doostan10b, Doostan11a, Yan12,Yang13,Peng14, Hampton14,Jones15,Jakeman15}. In this work, we adopt $\ell_1$-minimization to estimate the coefficients, solving the problem
\begin{equation}
\mathop{\arg\min}_{\bm c}\Vert \bm c\Vert_1\quad\mbox{subject to}\quad\Vert \bm u - \bm{\Psi c}\Vert_2 \leq \delta,
\label{eq:l1}
\end{equation}
where the vector $\bm c:=(c_1,\dots,c_P)$ contains PCE coefficients, while the vector $\bm u$ and the so-called {\it measurement matrix} $\bm\Psi$ contains function evaluations at realizations of the random input, $\bm\Xi$. Specifically, denoting the $i$th realization of $\bm\Xi$ as $\bm\xi^{(i)}$,
\begin{equation}
\bm u:=\left(u(\bm\xi^{(1)}),\dots,u(\bm\xi^{(N)})\right)^T;
\end{equation}
\begin{equation}
\label{eqn:psidef}
\bm \Psi(i,j):=\psi_j\left(\bm\xi^{(i)}\right).
\end{equation}
In \eqref{eq:l1}, $\delta$ is a tolerance parameter necessitated by the truncation error, $\epsilon_t(\bm{\Xi})$, to reduce the effect of overfitting. For this work, $\delta$ is identified via cross-validation~\cite{CrossValidation}.

It has been shown that as the number of samples, $N$, increases, the probability of accurately recovering PCE coefficients experiences a corresponding increase~\cite{Hampton14}. Often the deterministic solver of $u$ is computationally demanding, and it is expensive to compute realized $u(\bm\xi^{(i)})$. In order to deal with this situation, coefficient estimation based on gradient-enhanced PCE has received recent attention~\cite{Roderick10,Alekseev11,Li11,deBaar12,Lockwood13,Jakeman15}. Specifically, the gradient information utilized is
\begin{equation}
\bm u_{\partial}:=\left(\frac{\partial u}{\partial \Xi_1}\left(\bm\xi^{(1)}\right),\dots,\frac{\partial u}{\partial \Xi_d}\left(\bm\xi^{(1)}\right),\dots,\frac{\partial u}{\partial \Xi_1}\left(\bm\xi^{(N)}\right),\dots, \frac{\partial u}{\partial \Xi_d}\left(\bm\xi^{(N)}\right)\right)^T
\label{eqn:u_par}
\end{equation}
%\begin{equation}
%\bm u_{\partial}:=\left(\left.\frac{\partial u(\bm\Xi)}{\partial \Xi_1}\right|_{\bm\Xi=\bm\xi^{(1)}},\dots,\left.\frac{\partial u(\bm\Xi)}{\partial \Xi_d}\right|_{\bm\Xi=\bm\xi^{(1)}},\dots,\left.\frac{\partial u(\bm\Xi)}{\partial \Xi_1}\right|_{\bm\Xi=\bm\xi^{(N)}},\dots,\left.\frac{\partial u(\bm\Xi)}{\partial \Xi_d}\right|_{\bm\Xi=\bm\xi^{(N)}}\right)^T
%\label{eqn:u_par}
%\end{equation}
%
and
\begin{equation}
\bm\Psi_{\partial}((i-1)\cdot d+k,j):=\frac{\partial \psi_j}{\partial \Xi_k}(\bm\xi^{(i)}),\quad k=1,\dots,d.
\label{eqn:psi_par}
\end{equation}
With this gradient-enhancement, $\bm c$ is solved by minimizing $\|\bm{c}\|_2$, in least squares regression, or $\|\bm{c}\|_1$, in compressive sampling, subject to a constraint on
\begin{equation}
\label{eqn:grad_ell1}
\left\Vert \left(
\begin{aligned}
&\bm u\\
&\bm u_{\partial}
\end{aligned}
\right) - 
\left(
\begin{aligned}
&\bm \Psi\\
&\bm \Psi_{\partial}
\end{aligned}
\right)
\bm{W}\bm{c}\right\Vert_2,
\end{equation}
where $\bm{W}$ is a positive-diagonal matrix depending on the basis functions and their derivatives, which we shall specify in Section~\ref{subsec:gradient_analysis}. In this way, $N$ realizations of the random inputs $\bm{\Xi}$ provide an $N(d+1)\times P$ matrix, where the gradients $\partial u/\partial \Xi_k,~k=1,\dots,d,$ may be computed at each evaluation of $\bm{\Xi}$ from, e.g., direct or adjoint sensitivity equations~\cite{komkov86}, or automatic differentiation~\cite{Rall81}. We assume that $u(\bm\Xi)$ and its partial derivatives are square integrable with respect to the measure for $\bm{\Xi}$, and are represented in the appropriate basis such that the QoI and its derivatives correspond to identical coefficients.

In a related context, a gradient-enhanced sparse approximation based on $\ell_1$-minimization was proposed in~\cite{Tang13}, in which derivatives $\partial u/\partial \Xi_k,~k=1,\dots,d,$ are projected onto Legendre PCE, with a corresponding coefficient vector $\bm c_\partial$.

\subsection{Contribution of this work}
\label{subsec:contribution}

In this work, we investigate a gradient-enhanced $\ell_1$-minimization approach as seen in~\cite{Jakeman15}, in which the derivative information is empirically shown to improve the resolution of PCE coefficients. 

In Section~\ref{sec:theory}, a theoretical contribution concerning the Restricted Isometry Constant (RIC) is presented regarding recovery in $\ell_1$-minimization with derivative information based on Hermite PCE. The RIC is a constant associated with the measurement matrix that provides fruitful (probabilistic) bounds for recovery of solutions via $\ell_1$-minimization~\cite{Candes08c}. %%Alireza:, and hence providing a probabilistic bounds yields bounds for recovery as an extension. 
Additionally, under an appropriate normalization introduced by $\bm W$ in (\ref{eqn:grad_ell1}), it is guaranteed that null-space, measurement matrix column inner-products, and coherence related measures are almost-surely improved, implying that stability for solutions computed by (\ref{eq:l1}) are not reduced by including derivative information. These analyses are original to the authors' best knowledge, and provide a framework for analysis of recovery in other PCE bases. Also, though not considered here, the approach may be extended to least squares polynomial chaos regression~\cite{Hampton15}.

Three numerical experiments are used to demonstrate the gradient-enhanced $\ell_1$-minimization in Section~\ref{sec:example}, among which a plane Poiseuille flow with random boundaries is simulated, and the derivative information is approximated via an adjoint sensitivity method. The empirical results agree with the theoretical analysis, and suggest that the inclusion of derivative information can improve solution recovery at a lower overall computational cost.

The structure of the manuscript is as follows. In Section~\ref{sec:GEL1}, we state our problem and introduce the formulation of the gradient-enhanced $\ell_1$-minimization approach. In Section~\ref{sec:theory}, we present theoretical results concerning the stability and convergence of the gradient-enhanced $\ell_1$-minimization approach, in the context of Hermite PCE. In Section~\ref{sec:example}, we demonstrate our analysis empirically via three numerical examples: a manufactured PCE, an elliptic partial differential equation with random inputs, and a plane Poiseuille flow with random boundaries. Section~\ref{sec:proofs} presents the proofs to the results in Section~\ref{sec:theory}.

\section{Method Synopsis}
\label{sec:GEL1}
\subsection{Problem statement}
We use differential equations to model engineering systems on a domain $\mathcal{D}\in\mathbb{R}^D$, $D\in\{1,2,3\}$, {\color{black}in which} the uncertainty sources characterized by $\bm{\Xi}$ may be represented in one or many relevant parameters, e.g., boundary conditions and/or initial conditions. The solution $u$ is governed by the equations
\begin{equation}
\label{eqn:PDE_operator}
\begin{aligned}
&\mathcal{L}(\bm{x},t,\bm{\Xi};u(\bm{x}, t,\bm{\Xi}))=0,~~& &\bm{x}\in\mathcal{D},  \\
&\mathcal{I}(\bm{x},\bm{\Xi};u(\bm{x},0,\bm{\Xi}))=0,~~& &\bm{x}\in\mathcal{D}, \\
&\mathcal{B}(\bm{x},t,\bm{\Xi};u(\bm{x}, t,\bm{\Xi}))=0,~~& &\bm{x}\in\partial\mathcal{D}, 
\end{aligned}
\end{equation}
where $\mathcal{L},~\mathcal{I},$ and $\mathcal{B}$ are differential operators depending on the physics of the problem, the initial conditions, and the boundary conditions, respectively. Our objective is to approximate the QoI, here $u(\bm{x}_0,t_0,\bm{\Xi})$, for some fixed spatial location $\bm{x}_0$ and time $t_0$. Since we denote the realizations of the {\color{black}random inputs} by $\bm\xi^{(i)}$, the corresponding output is $u(\bm{x}_0,t_0,\bm{\xi}^{(i)})$. To reduce notation, we drop the reference to $\bm{x}_0$ and $t_0$, and simply write $u(\bm{\Xi})$ and $u(\bm{\xi}^{(i)})$.

\subsection{Polynomial chaos expansion (PCE)}
\label{subsec:PCE}
We rely on the PCE (\ref{eq:PCE}) to approximate the QoI, $u(\bm\Xi)$. For convenience, we assume that the input random variables, $\Xi_k$, are independent and identically distributed according to the probability density function $\rho_k$, and define $\{\psi_{i_k}(\Xi_k)\}$ to be the complete set of polynomials of degree $i_k\in\mathbb{N}\cup\{0\}$ orthogonal with respect to the weight function $\rho_k$ \cite{Askey85,Xiu02}. Hence, the multivariate orthonormal polynomials in $\bm{\Xi}$ are given by the products of the univariate orthonormal polynomials,
\begin{equation}
\label{eqn:PCProdDef}
\psi_{\bm{i}}(\bm{\Xi})=\mathop{\prod}\limits_{k=1}^d\psi_{i_k}(\Xi_k),
\end{equation}
where each $\bm i\in\{(i_1,\dots,i_d): i_k\in\mathbb{N}\cup\{0\}\}$ is a $d$-dimensional multi-index of non-negative integers. For computation, we truncate the expansion in (\ref{eq:PCE}) to the set of $P$ basis functions associated with the subspace of polynomials of total order not greater than $p$, that is $\sum_{k=1}^d i_k\le p$. For convenience, we also order these $P$ basis functions so that they are indexed by $\{1,\dots,P\}$, as in (\ref{eq:PCE}), where there should be no confusion in using either notation. The basis set $\{\psi_{j}(\bm\Xi)\}_{j=1}^P$ has cardinality
\begin{equation}
\label{eqn:P}
P = \frac{(d+p)!}{d!p!}.
\end{equation}
\subsection{$\ell_1$-minimization with gradient information}
We generate derivative information for the QoI, denoted by $\bm u_\partial$, and correspondingly evaluate the derivatives of ${\color{black}\psi_j(\bm\Xi)},~j=1,\dots,P{\color{black},}$ at the realizations $\bm{\xi}^{(i)},~i=1,\dots,N$, stored in a matrix $\bm\Psi_\partial$, as in (\ref{eqn:u_par}) and (\ref{eqn:psi_par}). For brevity, we define 
\begin{equation}
\tilde{\bm u}=\left(\begin{aligned}
&\bm u\\
&\bm u_{\partial}
\end{aligned}
\right)
\end{equation}
and
\begin{equation}
\tilde{\bm \Psi}=\left(
\begin{aligned}
&\bm \Psi\\
&\bm \Psi_{\partial}
\end{aligned}
\right)~,
\end{equation}
where $\tilde{\bm u}\in\mathbb{R}^{N(d+1)\times1}$, and $\tilde{\bm \Psi}\in\mathbb{R}^{N(d+1)\times P}$ is referred to as the \textit{gradient-enhanced measurement matrix}. Gradient-enhanced $\ell_1$-minimization solves the problem
\begin{equation}
\mathop{\arg\min}_{\bm c}\Vert \bm c\Vert_1\quad\mbox{subject to}\quad\left\Vert \tilde{\bm{u}} - 
\tilde{\bm{\Psi}}\bm{W}
\bm{c}\right\Vert_2 \leq \delta,
\label{eq:gel1}
\end{equation}
where $\delta$ generally differs from the choice in (\ref{eq:l1}) and $\tilde{\bm\Psi}$ is assumed to be normalized such that ${\color{black}\mathbb{E}\left[N^{-1}\tilde{\bm\Psi}^T\tilde{\bm\Psi}\right]= \bm{I}}$, the $P\times P$ identity matrix. Here, $\bm{W}$ is a positive-diagonal matrix, whose definition is deferred until Section~\ref{subsec:gradient_analysis}. We next begin a theoretical development to justify this approach.

\section{Theoretical Discussion}
\label{sec:theory}
We present results supporting the premise that the inclusion of derivative information does not reduce the stability of solutions recovered via (\ref{eq:gel1}) when compared to solutions recovered via (\ref{eq:l1}), i.e., in the absence of derivative information. We refer to these solutions and the methods to attain them using the adjectives gradient-enhanced and standard, respectively. We perform our analysis here with Hermite polynomials as they possess the convenient property that they and their derivatives are orthogonal with respect to the same measure, as by (5.5.10) of~\cite{Szego}. While we consider the Probabilists' polynomials here for exposition, the Physicists' polynomials would produce analogous results. Though we do not consider the details here, this analysis may be extended to the case of Laguerre or Jacobi polynomials where the derivative polynomials form an orthogonal system with respect to a measure that differs from the orthogonality measure, yet is still explicitly known, as by (5.1.15) and (4.21.7) of~\cite{Szego}, respectively.

First, we motivate and summarize results for standard $\ell_1$-minimization. Then we expand on these results in the case of gradient-enhanced $\ell_1$-minimization. The path of this analysis flows through the Restricted Isometry Constant (RIC)~\cite{Candes06a}, which is denoted by $\delta_s(\bm{\Phi})$ and is defined to be the smallest number satisfying
\begin{align}
\label{eq:ric_def}
(1-\delta_s(\bm{\Phi}))\|\bm{y}\|^2_2\le \|\bm{\Phi y}\|^2_2\le(1+\delta_s(\bm{\Phi}))\|\bm{y}\|^2_2.
\end{align}
Here, $\delta_s(\bm{\Phi})$ yields a uniform bound on the spectral radius of the submatrices of $\bm{\Phi}$ formed by selecting any $s$ columns of $\bm{\Phi}$. Often, the matrix being considered is clear from context, and we then shorten $\delta_s(\bm{\Phi})$ to $\delta_s$. Related to the RIC are restricted isometry properties that occur when the RIC reaches a small enough threshold, and guarantee that $\ell_1$-minimization with the given matrix is a stable computation. An example of such a restricted isometry property is given in Theorem~\ref{thm:RauhutWard} from~\cite{RauhutWard}. This theorem shows that if $\delta_{2s}<3/(4+\sqrt{6})$, where $s$ is dictated by the specific problem, then a stable recovery is assured.
\begin{thm}~\cite{RauhutWard}
\label{thm:RauhutWard}
Let $\bm{c}\in\mathbb{R}^P$ represent a solution we seek to approximate, and let $\hat{\bm{c}}$ be the solution to (\ref{eq:l1}). Let
\begin{align*}
\bm{\eta} &:= \bm{\Psi c}-\bm{u},
\end{align*}
denote the contribution from sources of error, and let $\epsilon$ from (\ref{eq:l1}), be chosen such that $\|\bm{\eta}\|_2^2<\epsilon$. If
\begin{align*}
\delta_{2s}(\bm{\Psi})&<\delta_{\star} := 3/(4+\sqrt{6})\approx 0.4652,
\end{align*}
then the following error estimates hold,
\begin{align*}
\|\bm{c}-\hat{\bm{c}}\|_2 &\le \frac{c_1}{\sqrt{s}}\mathop{\inf}\limits_{\|\bm{c}_s\|_0\le s}\|\bm{c}_s-\bm{c}\|_1 + c_2\epsilon;\\
\|\bm{c}-\hat{\bm{c}}\|_1 &\le c_3\mathop{\inf}\limits_{\|\bm{c}_s\|_0\le s}\|\bm{c}_s-\bm{c}\|_1 + c_4\epsilon\sqrt{s},
\end{align*}
 where $c_1,c_2,c_3,$ and $c_4$ depend only on $\delta_{2s}$, and $\|\cdot\|_0$ refers to the number of non-zero elements of the vector.
\end{thm}
We note that, related to the discussion in Section \ref{subsec:standard_analysis}, $\epsilon$ in Theorem \ref{thm:RauhutWard} may be selected so that $\|\bm{\eta}\|_2^2<\epsilon$ holds with high probability. Unfortunately, identifying the RIC for a given matrix requires a computation for every submatrix of $s$ columns, which is intractable in most situations of interest. As a practical alternative we instead choose to bound the RIC in a probabilistic sense, allowing us to identify a probability that the RIC is below a chosen threshold. In this way we can guarantee that a restricted isometry property, such as the one of Theorem~\ref{thm:RauhutWard}, holds with a certain probability. To do so, we introduce a definition of coherence, first considering the standard case, before expanding its definition to the gradient-enhanced case.

\subsection{Standard $\ell_1$-minimization analysis}
\label{subsec:standard_analysis}
We consider here an approach that uses arguments similar to those in~\cite{CandesPlan, Hampton14}. We note that those works did not proceed through the RIC as we do here. First, let $\mathcal{Q}$ be an arbitrary subset of the sample space for $\bm{\Xi}$, that is the values which $\bm{\Xi}$ can take. Here, $\mathcal{Q}$ is used to truncate the domain to one on which the basis functions, here Hermite polynomials, can be uniformly bounded. The coherence parameter for the standard approach~\cite{CandesPlan, Hampton14} is defined to be
\begin{align}
\label{eqn:std_coh_def}
 \mu_{\mathcal{Q}}:=\mathop{\sup}\limits_{k,\bm{\xi}\in\mathcal{Q}}|\psi_k(\bm{\xi})|^2_{2},
\end{align}
which for a precompact $\mathcal{Q}$ is guaranteed to be finite. An example of a $\mathcal{Q}$ suitable for use with Hermite polynomials, and used in~\cite{Hampton14, Hampton15}, is
\begin{align}
\label{eqn:trunc_def}
\mathcal{Q}:=\{\bm{\xi}:\|\bm{\xi}\|_2^2 \le (4+\epsilon_{p,d})p+2\},
\end{align}
where $\epsilon_{p,d}$ is a positive constant, which may be arbitrarily small but close to zero in an asymptotic analysis of the behavior of Hermite polynomials. We note that this truncation has not been analyzed when the number of samples is exponentially greater than the number of basis functions; however, this is not an issue here where the number of samples is typically less than or not substantially greater than the number of basis functions. The definition of coherence parameter as in (\ref{eqn:trunc_def}) leads to the following theorem, taken from Theorem 4.1 of~\cite{Hampton14}.
\begin{thm}
\label{thm:1dcoh}
For $d$-dimensional polynomials of order $p\ge 1$, the coherence in (\ref{eqn:std_coh_def}) is bounded by
\begin{align}
 \label{eqn:std_coh_bd}
\mu_{\mathcal{Q}}\le C_0\cdot C_1^{p},
\end{align}
where $C_0$ and $C_1$ are modest constants depending on $d,p,\epsilon_{p,d}$. As $p/d\rightarrow\infty$, $C_0$ decreases to $1$, and $C_1$  decreases to a limit of $\exp(2-\log(2))\approx 3.7$.
\end{thm}
This shows an exponential dependence of the coherence parameter on $p$.  Let $\bm{X}_k$ denote a row vector consisting only of basis polynomial evaluations at $\bm{\xi}^{(k)}$. We bound the RIC for the matrix $\bm{\Psi}$ defined as in (\ref{eqn:psidef}).
\begin{thm}
\label{thm:1}
For any chosen $\mathcal{Q}$, we may bound the RIC in a probabilistic sense by
\begin{align*}
\mathbb{P}(\delta_s< t) \ge \mathbb{P}(\mathcal{Q})^N-\exp\left(-C_{\mathcal{Q}}\frac{Nt}{s\mu_{\mathcal{Q}}}+s+\log(2s)+s\log(P/s)\right).
\end{align*}
\end{thm}
\begin{remark}
\label{rem:truncation}
We note that we do not generally require arbitrarily small $\delta_s$, as is seen in Theorem~\ref{thm:RauhutWard}. The constant $C_{\mathcal{Q}}$ scales with 
\begin{align}
\label{eqn:trunc_err}
\epsilon_{\mathcal{Q}}:=\left\|\mathbb{E}\left(\bm{X}^T\bm{X}|\bm\xi\in\mathcal{Q}\right)-\bm{I}\right\|_2,
\end{align}
which is a bias that is negligible in practical contexts for $\mathcal{Q}$ as in (\ref{eqn:trunc_def})~\cite{Hampton14}. Truncating in this way, neither $C_{\mathcal{Q}}$ nor $\mathbb{P}(\mathcal{Q})$ are problematic in practice.
\end{remark}
\begin{remark}
\label{rem:ell2}
While our primary focus here is $\ell_1$-recovery, the RIC corresponding to $s=P$ is useful for analyzing the stability of a least squares solution, and so this result is also applicable to $\ell_2$-minimization. For ways in which this parameter may bound error from solutions computed via $\ell_2$-minimization we point the interested reader to~\cite{CDL13, Hampton15}. In this case, a slight adjustment to the proof gives the bound
\begin{align*}
\mathbb{P}(\delta_P< t) \ge \mathbb{P}(\mathcal{Q})^N-2P\exp\left(-C_{\mathcal{Q}}\frac{Nt}{P\mu_{\mathcal{Q}}}\right).
\end{align*}
\end{remark}
The following corollary, which follows from a rearrangement of the result of Theorem~\ref{thm:1}, highlights the relationship between several quantities.
\begin{cor}
\label{cor:sample_bound}
To insure that $\delta_s < \delta_{\star}$ with probability $p_{\star}$, it is sufficient to take $N_{\star}$ satisfying
\begin{align*}
N_{\star}\delta_{\star} \ge \frac{s\mu_{\mathcal{Q}}}{C_{\mathcal{Q}}}\left[s+\log(2s)+s\log(P/s) - \log\left(\mathbb{P}(\mathcal{Q})^{N_{\star}}-p_{\star}\right)\right].
\end{align*}
\end{cor}
For example,  using $\delta_{\star}$ as in Theorem~\ref{thm:RauhutWard}, gives a guarantee for stability of solutions to (\ref{eq:l1}) when the number of samples $N$ satisfies
\begin{align*}
N \ge \frac{(4+\sqrt{6})s\mu_{\mathcal{Q}}}{3C_{\mathcal{Q}}}\left[s+\log(2s)+s\log(P/s) - \log\left(\mathbb{P}(\mathcal{Q})^{N}-p_{\star}\right)\right].
\end{align*}
We note that $N$ appears on both the left and right sides of the equation. This arises from the truncation, which has a technical issue when $N$ is exponentially larger than $P$; specifically, the probability that at least one sample had fallen in $\mathcal{Q}^c$ becomes large, while the analysis relies on this event being rare. As $N$ is very often smaller than $P$, and rarely chosen to be exponentially larger than $P$, this issue is not of practical concern. Next, we extend these results to the case where gradient information is included.
\subsection{Gradient-enhanced $\ell_1$-minimization analysis}
\label{subsec:gradient_analysis}
In the previous case the rows of $\bm{\Psi}$ were independent, while $\tilde{\bm{\Psi}}$ has a more subtle independence structure, as only the sets of rows associated with the independent samples are independent. Related to this point, we let the $(d+1)\times P$ block of independent information related with the $k$th sample be given by $\bm{X}_k$, with a generic realization given by $\bm{X}$. Specifically,
\begin{align*}
\bm{X}(i,j) &= \frac{\partial\psi_j}{\partial\Xi_i}({\color{black}\bm{\xi}}), \quad i=1,\dots,d;\\
\bm{X}(d+1,j) &= \psi_j({\color{black}\bm{\xi}}).
\end{align*}
That is, the last row corresponds to the realizations of the function, while the first $d$ rows correspond to the derivative information. We note that there is no effect in the computations considered here by rearranging the rows of the matrix $\tilde{\bm{\Psi}}$. The RIC will necessarily be larger if the norms of the matrix columns are different. Adjusting for this can be done by adjusting the basis functions themselves. In the case of standard $\ell_1$-minimization, we use orthonormal polynomials. Here, we will wish to include derivatives of those polynomials as well, requiring a different normalization.  We use the Probabilists' Hermite polynomials, and the following lemma is used to identify this normalization.
\begin{lem}
\label{lem:hermite_normalize}
For $d$-dimensional orthonormal Probabilists' Hermite polynomials $\psi_{\bm{i}}$ and $\psi_{\bm{j}}$ with order $i_k,j_k$ in dimension $k$,
\begin{align}
\label{eqn:herm_norm}
\mathbb{E}\left(\psi_{\bm{i}}(\bm{\Xi})\psi_{\bm{j}}(\bm{\Xi}) + \mathop{\sum}\limits_{k=1}^d\frac{\partial\psi_{\bm{i}}}{\partial\Xi_k}(\bm{\Xi})\frac{\partial\psi_{\bm{j}}}{\partial\Xi_k}(\bm{\Xi})\right)  = \delta_{\bm{i},\bm{j}}\left(1+\mathop{\sum}\limits_{k=1}^d i_k\right),
\end{align}
where $\delta_{\bm{i},\bm{j}}$ is the Kronecker Delta.
\end{lem}
This suggests a different normalization of the basis functions to enforce that columns of the gradient-enhanced measurement matrix %%, $\tilde{\bm{\psi}}$,
 have the same expected $\ell_2$-norm. We refer to this normalization as {\it gradient-normalization}. Specifically, we multiply the orthonormal basis function $\psi_{\bm{i}}$ by 
\begin{align}
\label{eqn:weight_def}
\tilde{w}_{\bm{i}}:=\left(1+\mathop{\sum}\limits_{k=1}^d i_k\right)^{-1/2},
\end{align}
to gradient-normalize. We note that it is these weights that define the $\bm{W}$ in (\ref{eqn:grad_ell1}) and \eqref{eq:gel1}. In this work, we assume that when derivative information is included, that those basis functions are gradient-normalized, and that when derivative information is not included, that those basis functions are orthonormal, which we refer to as standard-normalization. This insures that the expected norms for the columns of the sampled matrices are consistent in both cases. 

Recalling that $N$ denotes the number of samples used, our analysis focuses on the Gramian matrix
\begin{align}
\label{eqn:m_def}
\bm{M} := \frac{1}{N}\mathop{\sum}\limits_{k=1}^{N}\bm{X}_k^T\bm{X}_k.
\end{align}
Here, $\bm{M} = N^{-1}\bm{\Psi}^T\bm{\Psi}$ for the standard approach, while $\bm{M}=N^{-1}\tilde{\bm{\Psi}}^T\tilde{\bm{\Psi}}$ for the gradient-enhanced approach. We now present some summarized results for the standard approach that will be compared to the results presented for the gradient-enhanced case.
To analyze the spectrum of $\bm{M}$ in the case of gradient-enhanced $\ell_1$-minimization, we use the following definition, which generalizes the $\ell_1$-coherence as studied in~\cite{CandesPlan, Hampton14, RauhutWard} and defined in (\ref{eqn:std_coh_def}). %%from the case of independent rows to the case of independent sets of rows.
 Let $\mathcal{Q}$ be an arbitrary subset of $\Omega$, which we use to truncate the sample space to insure a uniformly bounded polynomial system, e.g. as in~\cite{CandesPlan, Hampton14, Hampton15}, and let
\begin{align}
\label{eqn:matrix_coh_def}
 \beta_{\mathcal{Q}}:=\mathop{\sup}\limits_{k,\bm{\xi}\in\mathcal{Q}}\|\bm{X}(:,k)\|^2_{2}.
\end{align}
This parameter is a generalization of $\mu_{\mathcal{Q}}$ in the case where rows are not independent, but sets of rows are. Note that in the case that $\bm{X}$ is a row vector, then this definition reduces to (\ref{eqn:std_coh_def}). Specifically, the results of Section~\ref{subsec:standard_analysis} all hold when substituted for this parameter, which we highlight as a theorem.
\begin{thm}
\label{thm:mu_beta}
The theorems of Section~\ref{subsec:standard_analysis} hold for the gradient-enhanced case when $\mu_{\mathcal{Q}}$ is replaced by $\beta_{\mathcal{Q}}$.
\end{thm}
We conclude our analysis with three results which demonstrate that the inclusion of derivative information, coupled with gradient-normalization, does not reduce stability over the corresponding approach without derivative information. 

{\color{black}The first result} is an inequality concerning the coherence parameter defined in (\ref{eqn:std_coh_def}) and (\ref{eqn:matrix_coh_def}), showing that including derivative information does not require any weakening of the bounds of Section~\ref{subsec:standard_analysis} in the case without derivative information. This inequality then directly applies to all of the theorems in Section~\ref{subsec:standard_analysis}.

The second result is a direct null-space comparison of the two different matrices, which is known to be fundamental for recovery of exactly-sparse solutions~\cite{Candes06a,Juditsky11a,Juditsky11b}. Specifically, as $\|\tilde{\bm{\Psi}}\bm{c}-\tilde{\bm{u}}\|<\delta$ (or $\|\bm{\Psi c}-\bm{u}\|<\delta$) is enforced, the difference between potential solutions is close to an element of the null-space of $\tilde{\bm{\Psi}}$ (or  $\bm\Psi$), so reducing the dimension of the null-space correspondingly reduces the space of potential solutions.

The third result concerns a bound on the inner-products of columns of the measurement-matrix. This is related to the RIC in that if the inner-product between several pairs of columns is of large absolute value then a linear combination of those columns will have small norm, resulting in a larger RIC. Similarly, if those inner-products are of small absolute value, then no linear combination will have a small norm. An analogous observation may be made regarding a linear combination of columns having a much larger norm. For this reason it is beneficial if the inner-product between columns is of small absolute value.
\begin{thm}
\label{thm:2}
Let $\tilde{\bm{\Psi}}$ be a realized measurement matrix with derivative information that is gradient-normalized. Similarly, let $\bm{\Psi}$ be a realized measurement matrix with standard-normalization and no derivative information. 

Assume that $\bm{\Psi}$ and $\tilde{\bm{\Psi}}$ are formed from the same realized input samples, $\{\bm{\xi}^{(i)}\}_{i=1}^N$, so that up to row weighting, $\bm{\Psi}$ is a sub-matrix of $\tilde{\bm{\Psi}}$. Then the following statements related to the recovery of solutions via $\ell_1$-minimization hold.

{\bf R1.} Using the definition in (\ref{eqn:matrix_coh_def}) for the two different approaches,
\begin{align*}
\beta_{\mathcal{Q}}(\tilde{\bm{\Psi}}) \le \mu_{\mathcal{Q}}(\bm{\Psi}),
\end{align*}
and this inequality is almost-surely strict.

{\bf R2.} If $\mathcal{N}(\cdot)$ represents the null-space, then $\mathcal{N}(\tilde{\bm{\Psi}})\subset\mathcal{N}(\bm{\Psi})$, and this is almost-surely a strict subset when $\bm{\Psi}$ is undersampled. Specifically, it almost-surely holds that $\dim(\mathcal{N}(\tilde{\bm{\Psi}}))=\max\{0,P-(d+1)N\}$ while $\dim(\mathcal{N}(\bm{\Psi}))=\max\{0,P-N\}$.

{\bf R3.} If subscripts of matrices correspond to the columns, and $i_k$ denotes the order of the basis polynomial in the $\bm{i}$th column in the $k$th dimension, then the associated inner product of columns is bounded by,
 \begin{align}
\mathop{\sup}\limits_{\bm{i}\ne \bm{j}}|(\tilde{\bm{\Psi}}_{\bm{i}},\tilde{\bm{\Psi}}_{\bm{j}})|&\le \mathop{\sup}\limits_{\bm{i}\ne \bm{j}}\frac{|(\bm{\Psi_i},\bm{\Psi_j})|(1 + \sum_{k=1}^d\sqrt{i_kj_k})}{\sqrt{(1+\sum_{k=1}^di_k)(1+\sum_{k=1}^dj_k)}}\le \mathop{\sup}\limits_{\bm{i}\ne \bm{j}}|(\bm{\Psi_i},\bm{\Psi_j})|.
\end{align}
\end{thm}
\begin{remark}
\label{rem:partial_enhancement}
The theorem is presented for full gradient information to ease presentation, but the three points generalize to the case that derivative information is included for a fraction of samples. Specifically, (1) holds with an appropriate adjustment to the dimensionality, (2) holds with an adjustment to the basis dependent multiplicative constant, and (3) holds if $\tilde{\bm{\Psi}}$ has derivative information for only a few samples. In summary adding derivative information for even a percentage of the samples leads to bounds as in Theorem~\ref{thm:2}.
\end{remark}
\subsection{A note on potentially contrasting solutions}
\label{subsec:loss_remark}
It is of practical importance to note what functions are recovered in an asymptotic sense by the gradient-enhanced and standard approaches, as they may differ significantly. The gradient-enhanced method gives $\hat{u}$ approximating $u$ in a Sobolev type loss function,
\begin{align}
\label{eq:loss_def}
L(\hat{u},u)&:=\|\hat{u}-u\|^2_{\ell_2(\bm{\Xi},N)}+\mathop{\sum}\limits_{k=1}^d\left\|\frac{\partial{(\hat{u}-u)}}{{\partial{\color{black}\Xi_k}}}\right\|^2_{\ell_2(\bm{\Xi},N)},
\end{align}
where the $\ell_2(\bm{\Xi},N)$ indicates the discrete $\ell_2$ norm using $N$ evaluations drawn from realizations of $\bm{\Xi}$, here {\color{black}$(\bm{\xi}^{(1)},\dots,\bm{\xi}^{(N)})$}. This norm is also normalized by $N^{-1}$ so that as $N$ goes to infinity this norm tends to $\mathcal{L}_2(\bm{\Xi})$, the standard $\mathcal{L}_2$ norm associated with the distribution of $\bm{\Xi}$. Specifically (\ref{eq:gel1}) guarantees that $NL(u,\hat{u})<\delta$.  In contrast, without derivative information using standard-normalization and producing a solution via (\ref{eq:l1}), gives $\hat{u}$ approximating $u$ such that $N\|\hat{u}-u\|^2_{\ell_2(\bm{\Xi},N)}<\delta$, that is the partial derivatives of the approximation need not be approximated by those of the target function. {\color{black}As these loss functions differ, so too does the limiting solution produced by each method for a finite expansion order $p$. However, the sequence of approximations given by the two loss functions and by using increasing $p$ (and accordingly $N$) will converge to $u$ in the $\mathcal{L}_2(\bm{\Xi})$ sense as $p,N\rightarrow\infty$. }

%As a toy example to highlight this difference, consider the one dimensional case where we seek to approximate the function {\color{black}$u(\Xi)=(a_1 \Xi - a_0)^2$}, where $\Xi$ follows a standard normal distribution. We will use an affine approximation $\hat{c}_0 + \hat{c}_1 \Xi${\color{black}, which is  a 1st order Hermite polynomial expansion. We} would like to compare the limiting approximations which is identified in this case by explicit computation. If we do not utilize derivative information, the limiting solution is 
%%
%\begin{align}
%\hat{u}_0(\Xi):=a_0^2+a_1^2 - 2a_0a_1\Xi.
%\end{align}
%%
%In contrast, if we utilize derivative information, (\ref{eq:loss_def}) implies that the limiting solution is 
%%
%\begin{align}
%{\color{red}\hat{u}_{1}(\Xi):=a_0^2+a_1^2-4a_0a_1\Xi.}
%\end{align}
%%
%{\color{red}We note that $\hat{u}_0-\hat{u}_{1}=2a_0a_1\Xi$. It follows that $\|\hat{u}_0-\hat{u}_{1}\|_{\mathcal{L}_2(\bm{\Xi})}=2a_0a_1$, which may be large, even when considering relative error using $\hat{u}_0$, $\hat{u}_{1}$ or $u$.} As a result, standard $\ell_1$-minimization seeks approximations which best match desired function values, while gradient-enhanced $\ell_1$-minimization also insures an approximation of the derivatives of the desired function, though at the potential cost of less accurate function values.
%
\section{Numerical Results}
\label{sec:example}
In this section, we empirically demonstrate the gradient-enhanced $\ell_1$-minimization approach via three numerical examples: a manufactured PC{\color{black}E}; an elliptic PDE with stochastic coefficient; and a {\color{black}plane Poiseuille flow with random boundaries}. To compare the standard and gradient-enhanced $\ell_1$-minimization solutions, we define an equivalent sample size $\tilde{N}$,
\begin{equation}
\label{eqn:tildeN}
\tilde{N} := N_e + \nu N_g,
\end{equation}
which accounts for the added cost of computing the derivative information. In (\ref{eqn:tildeN}), $N_e$ is the number of samples without derivative information, $N_g$ is the number of samples with derivatives (along all $d$ directions), and $\nu$ is a positive parameter depending on the problem at hand and the approach employed to compute the derivatives. For the example of Section~\ref{subsec:PDE}, the cost of generating $d$ derivatives of the QoI, obtained by the adjoint sensitivity method, is roughly the same as that of evaluating the QoI, thus implying that $\nu=2$. %We highlight that the equivalent sample size model \eqref{eqn:tildeN} has to be adjusted for a given problem. 
For transient problems for which the cost of solving the adjoint equations for derivative calculations may be considerably more than that of a single QoI evaluation, then $\nu>2$. Nevertheless, we here present all cost comparisons in terms of the number of equivalent sample size $\tilde{N}$ in \eqref{eqn:tildeN}, for choices of $\nu$ that we shall specify. For simulations based on standard $\ell_1$-minimization, we set $\tilde{N}=N_e$. Additionally, for the interest of convenience, we ignore the cost of solving the $\ell_1$-minimization problems in \eqref{eq:l1} and \eqref{eq:gel1}. This is a valid assumption as often that cost is negligible relative to the cost of evaluating the QoI or its derivatives. For terminology, if $X\%$ of samples used in the computation of the solution contain derivative information, then we say that method is $X\%$ gradient-enhanced. In this way, the standard approach without derivatives would give a solution that is 0\% gradient-enhanced, denoted as standard in the subsequent figures. Similarly, if all samples include derivative information, the associated solution would be 100\% gradient-enhanced.

\subsection{Case I: A manufactured PCE}

First, we consider the reconstruction of a manufactured PCE, in which the sparsity and the entries of the coefficient vector $\bm c$ are {\it a priori} prescribed. Specifically, we set the dimension of the expansion to $d=25$ and use a $p=3$ order PCE (hence $P=3276$ basis functions) to manufacture the QoI $u(\bm\Xi)$. To generate $\bm c$, we first draw its $P$ entries independently from the standard Gaussian distribution. We then retain $|\mathcal{C}|\in\{50,150\}$ coefficients with largest magnitude and set the rest to zero. This gives a randomized sparsity support. Finally, the realizations of $u(\bm\Xi)$ and its derivative with respect to $\Xi_k$, $k=1,\dots,d$, are generated by 

\begin{equation}
\label{eqn:manufactured_u}
u({\color{black}\bm\xi^{(i)}}) = \sum_{j=1}^{P}c_j\psi_j(\bm\xi^{(i)})~
\end{equation}
and
\begin{equation}
\label{eqn:manufactured_upar}
%u_{\partial_k}(\bm\xi^{(i)}) = \sum_{j=1}^{P}c^0_j\frac{\partial\psi_j(\bm\xi^{(i)})}{\partial\Xi_k}~, k=1,\dots,d~,
\color{black}
u_{\partial_k}(\bm\xi^{(i)}) := \sum_{j=1}^{P}c_j\frac{\partial\psi_j}{\partial\Xi_k}(\bm\xi^{(i)}), 
\end{equation}
respectively. %, where $u_{\partial_k}(\bm\xi^{(i)})$ is the derivative of $u(\bm\xi^{(i)})$ with respect to ${\color{red}\Xi_k}$, i.e.,
%%
%\begin{equation}
%%u_{\partial_k}(\bm\xi^{(i)}):=\frac{\partial u(\bm\xi^{(i)})}{\partial \Xi_k}~.
%\color{red}
%u_{\partial_k}(\bm\xi^{(i)}):=\left.\frac{\partial u(\bm\Xi)}{\partial \Xi_k}\right|_{\bm\Xi=\bm\xi^{(i)}}~.
%\end{equation}
%%
We then approximate the PCE coefficients $\bm c$ via (\ref{eq:l1}) and (\ref{eq:gel1}) from these generated data.

\subsubsection{Results}
Beginning with the $|\mathcal{C}|=50$ case, we seek to recover the manufactured $\bm{c}$. If the computed solution, denoted by $\hat{\bm{c}}$, has a relative root-mean-square-error (RRMSE) below $0.01\%$, then we call it a successful recovery of $\bm{c}$. 

\begin{figure}[H]
\centering
\subfloat[]{\label{fig:PCE_all_S150}\includegraphics[width=0.48\textwidth]{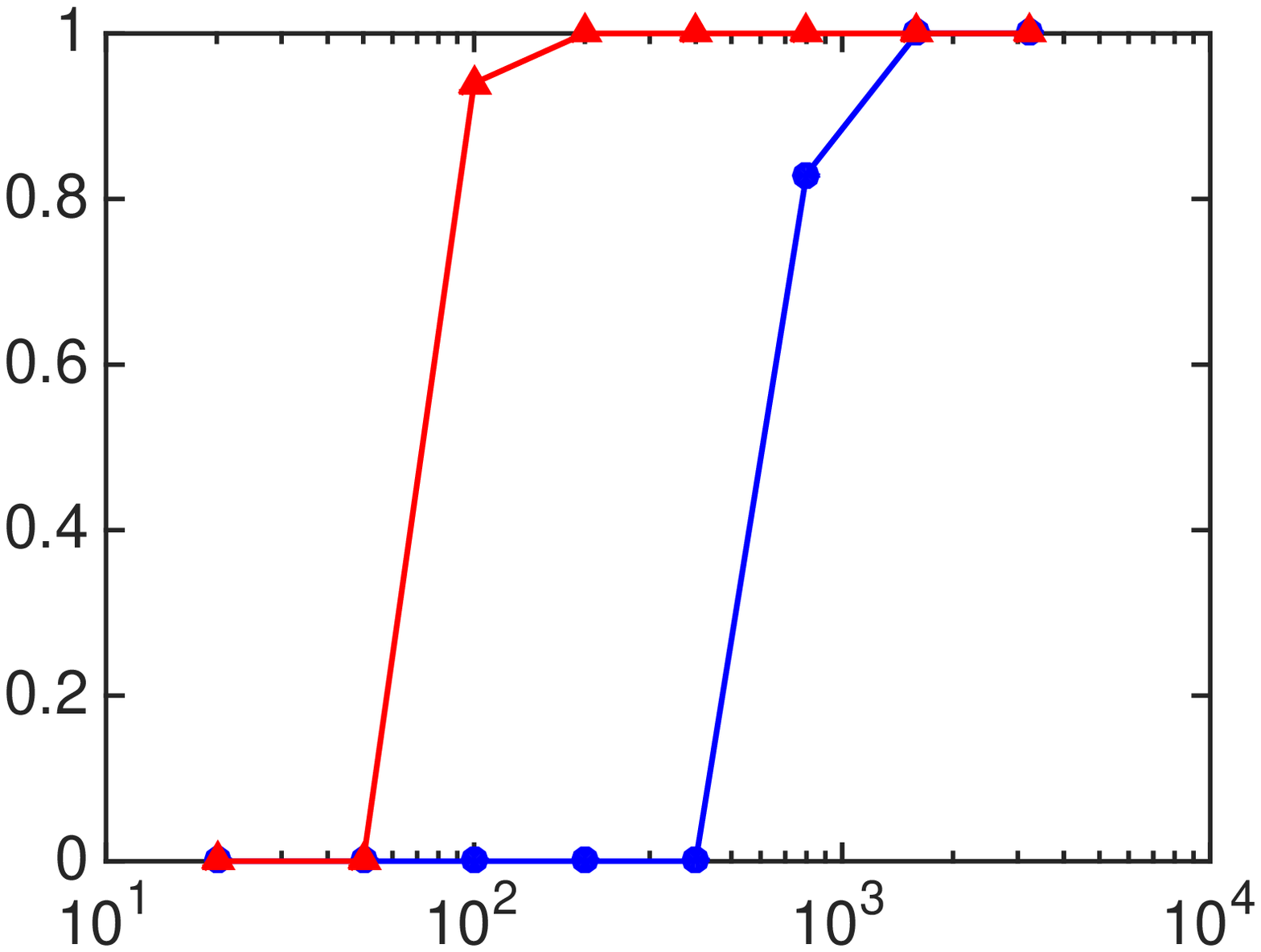}}
\put(-195,45){ \begin {sideways} {\scriptsize \fontfamily{phv}\selectfont Recovery probability} \end{sideways}}
\put(-95,0){{\scriptsize $\tilde{N}$}}
\hspace{.2cm}
\subfloat[]{\label{fig:PCE/PCE_20_S150}\includegraphics[width=0.48\textwidth]{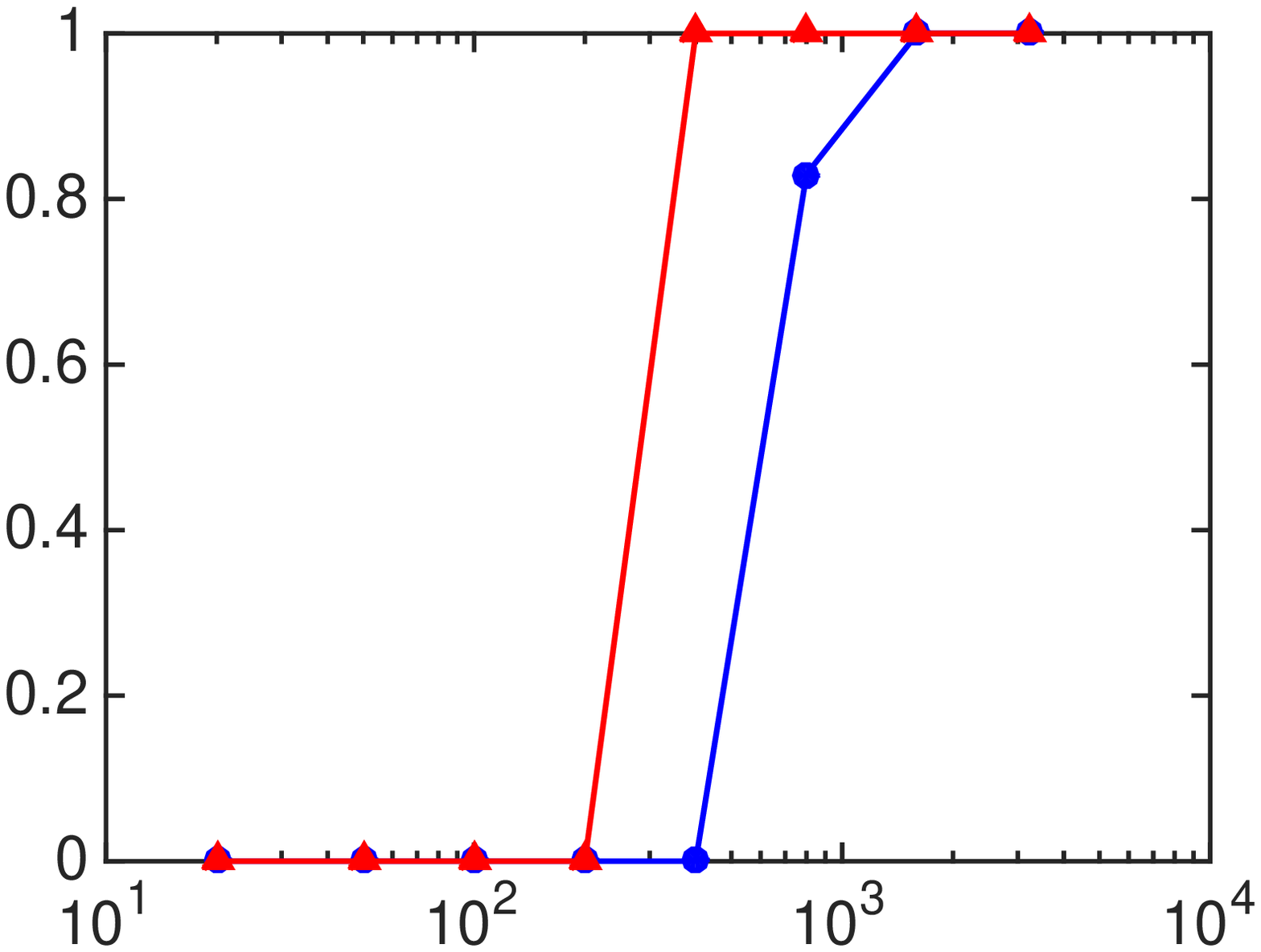}}
\put(-195,45){ \begin {sideways} {\scriptsize \fontfamily{phv}\selectfont Recovery probability} \end{sideways}}
\put(-95,0){{\scriptsize $\tilde{N}$}}
\caption{Probability of successful recovery of manufactures PCE with sparsity $|\mathcal{C}|=150$ via gradient-enhanced and standard $\ell_1$-minimization. (a) 100\% vs. 0\% gradient-enhanced.  (b) 20\% vs. 0\% gradient-enhanced. (\usebox{\LegendeRDU}~ gradient-enhanced, \usebox{\LegendeBC}~ standard)}
\label{fig:PCE_S150}
\end{figure}

\begin{figure}[H]
\centering
\subfloat[]{\label{fig:PCE_all_S50_noisy}\includegraphics[width=0.48\textwidth]{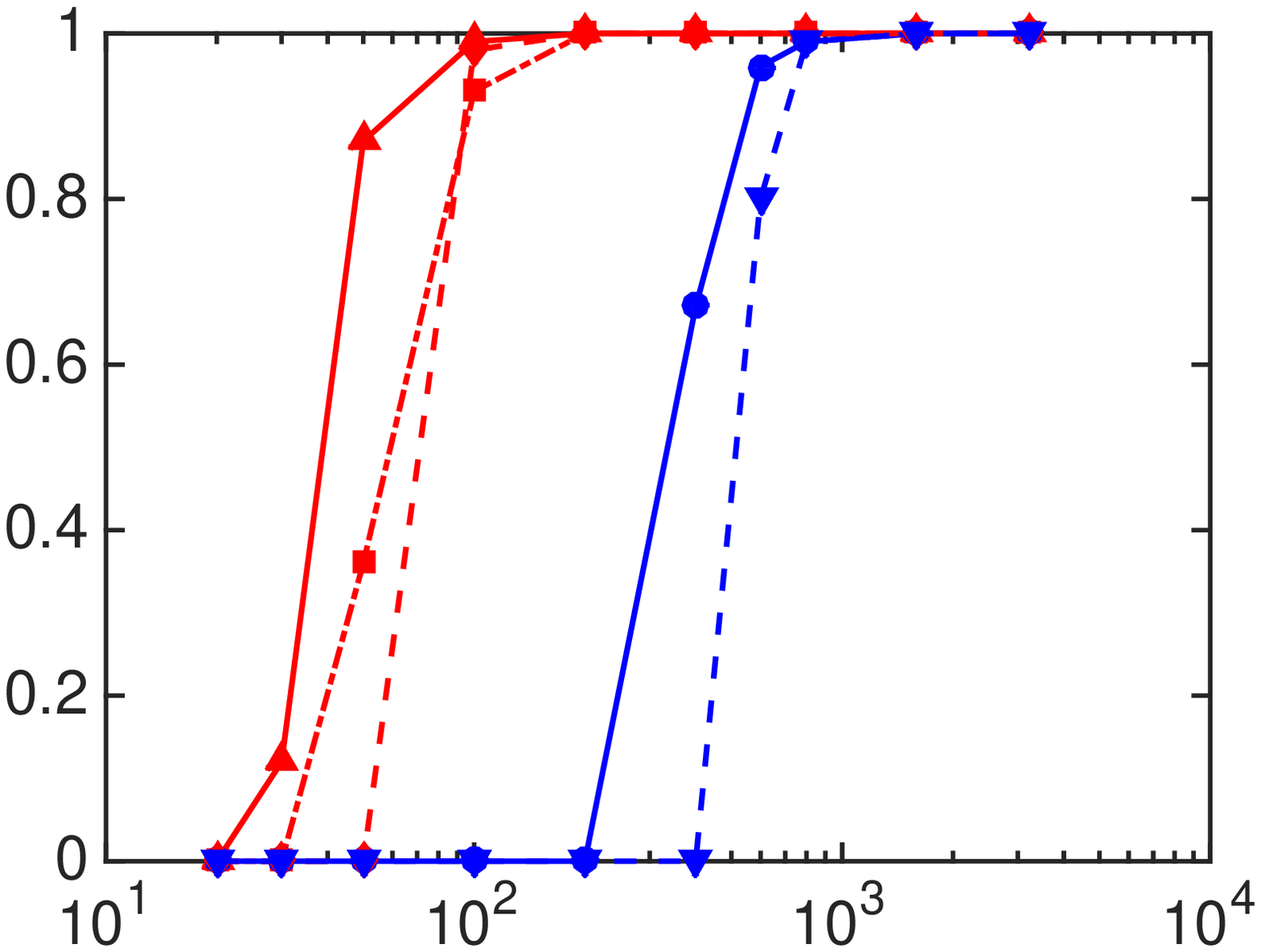}}
\put(-195,45){ \begin {sideways} {\scriptsize \fontfamily{phv}\selectfont Recovery probability} \end{sideways}}
\put(-95,0){{\scriptsize $\tilde{N}$}}
\hspace{.2cm}
\subfloat[]{\label{fig:PCE/PCE_partial_S50_noisy}\includegraphics[width=0.48\textwidth]{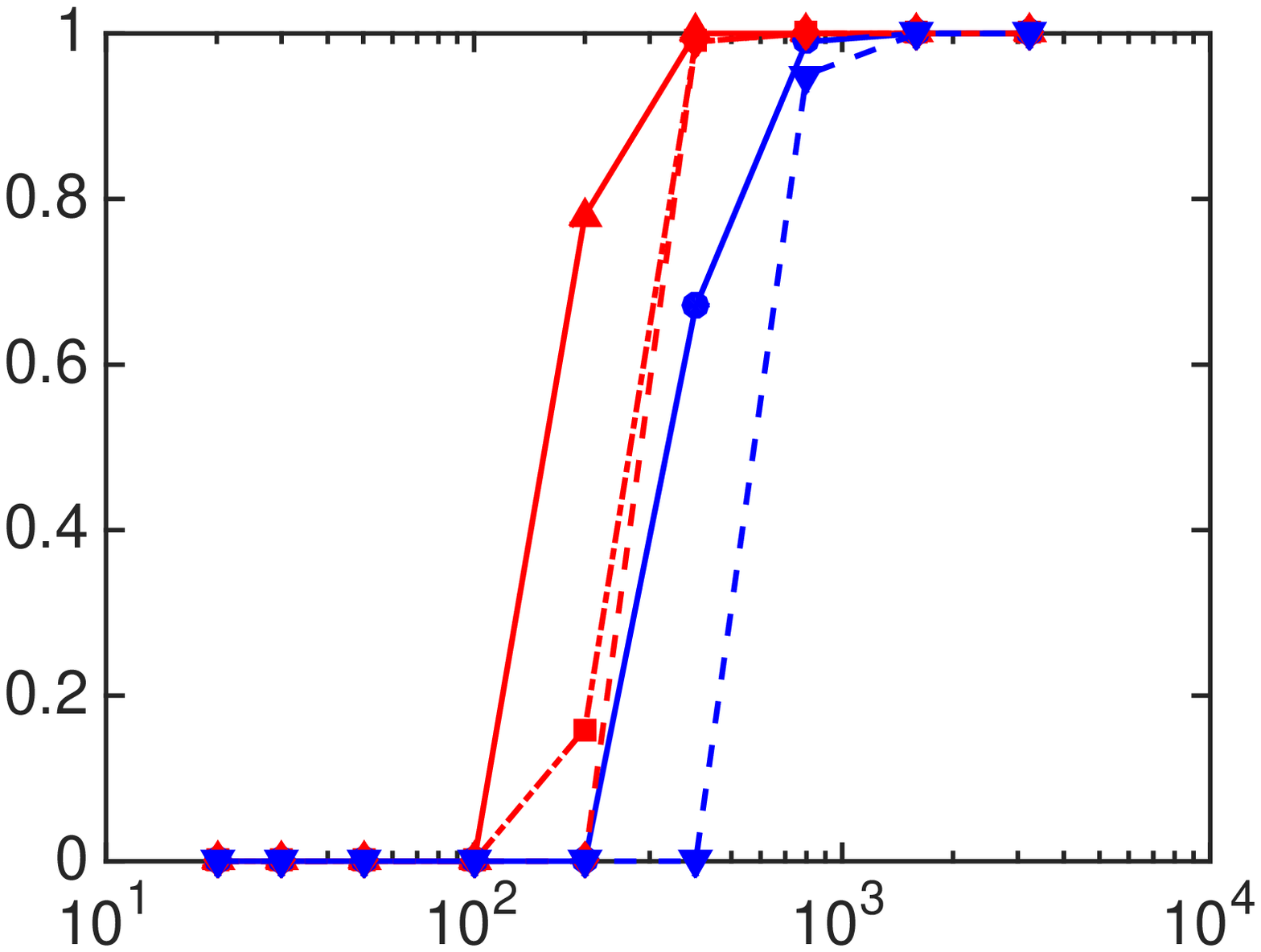}}
\put(-195,45){ \begin {sideways} {\scriptsize \fontfamily{phv}\selectfont Recovery probability} \end{sideways}}
\put(-95,0){{\scriptsize $\tilde{N}$}}
\caption{Probability of successful recovery of gradient-enhanced and standard $\ell_1$-minimization for the manufactured PCE case with sparsity $|\mathcal{C}|=50$. (a) 100\% gradient-enhanced. (b) 20\% gradient-enhanced (\usebox{\LegendeRDU}~gradient-enhanced, \usebox{\LegendeRSDd}~gradient-enhanced with noisy $\bm u_{\partial}$ only,  \usebox{\LegendeRDD}~gradient-enhanced with both noisy $\bm u$ and $\bm u_{\partial}$, \usebox{\LegendeBC}~standard,  \usebox{\LegendeBDT}~standard with noisy $\bm u$)}
\label{fig:PCE_S50_noisy}
\end{figure}

{\color{black}We first consider the case where evaluations of $u(\bm\Xi)$ and its derivatives are exact, i.e., noise free. In Fig.~\ref{fig:PCE_S50}, we compare the probability of successful recovery for gradient-enhanced and standard approaches, using $100$ independent replications for each $\tilde N$. To set $\tilde N$, we pretend the cost of evaluating $d$ derivatives of $u(\bm\Xi)$ is the same as that of evaluating $u(\bm\Xi)$, and therefore we set $\nu=2$.
\begin{figure}[H]
\centering
\subfloat[]{\label{fig:PCE_all_S50}\includegraphics[width=0.48\textwidth]{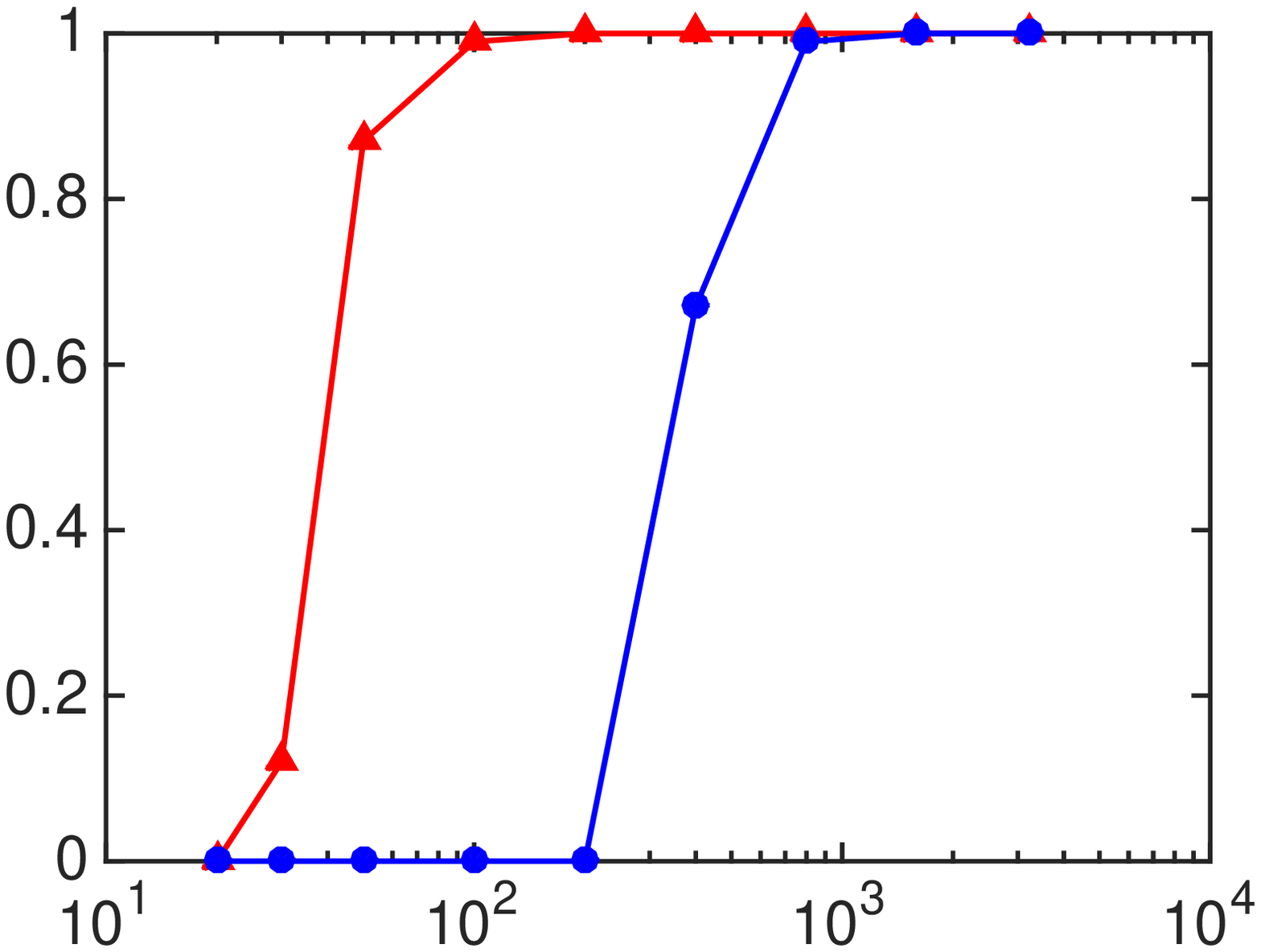}}
\put(-195,45){ \begin {sideways} {\scriptsize \fontfamily{phv}\selectfont Recovery probability} \end{sideways}}
\put(-95,0){{\scriptsize $\tilde{N}$}}
\hspace{.2cm}
\subfloat[]{\label{fig:PCE/PCE_partial_S50}\includegraphics[width=0.48\textwidth]{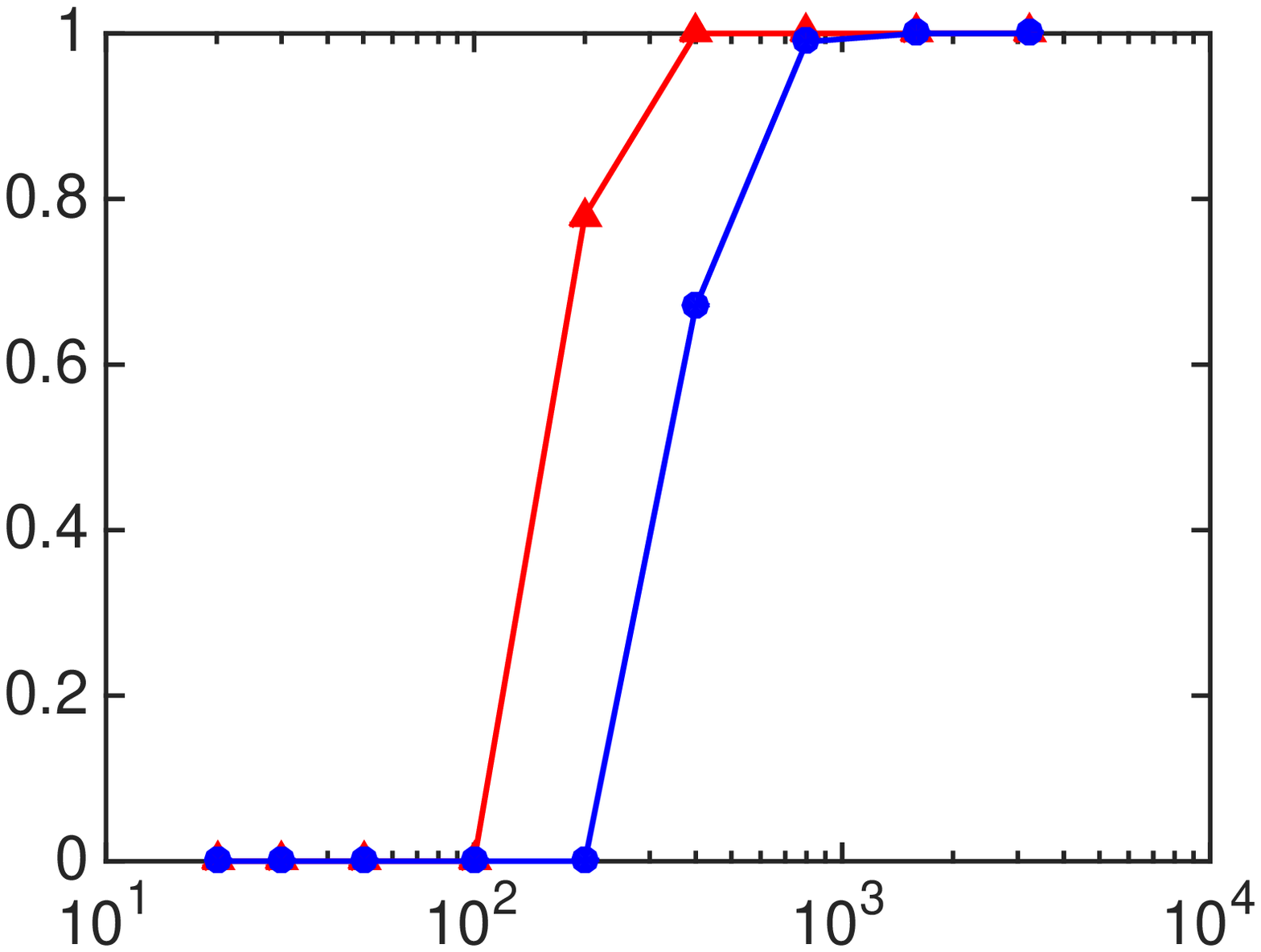}}
\put(-195,45){ \begin {sideways} {\scriptsize \fontfamily{phv}\selectfont Recovery probability} \end{sideways}}
\put(-95,0){{\scriptsize $\tilde{N}$}}
\caption{Probability of successful recovery of manufactures PCE with sparsity $|\mathcal{C}|=50$ via gradient-enhanced and standard $\ell_1$-minimization. (a) 100\% vs. 0\% gradient-enhanced.  (b) 20\% vs. 0\% gradient-enhanced. (\usebox{\LegendeRDU}~ gradient-enhanced, \usebox{\LegendeBC}~ standard)}
\label{fig:PCE_S50}
\end{figure}
In Fig.~\ref{fig:PCE_all_S50}, we see that 100\% gradient-enhanced $\ell_1$-minimization helps in reducing the computational effort to recover $\bm c$, while Fig.~\ref{fig:PCE/PCE_partial_S50}, demonstrates a notable, but less, improvement for 20\% gradient-enhanced $\ell_1$-minimization. The figure suggests that adding derivative information is a cost-effective means to increase the probability of successfully recovering $\bm c$ at low sample sizes $\tilde N$.

Fig.~\ref{fig:PCE_S150} shows similar results for sparsity $|\mathcal{C}|=150$. Since $|\mathcal{C}|$ is larger in this case, both standard and gradient-enhanced methods require more samples for recovery with a given success probability. This is consistent with the sampling rate presented in Corollary \ref{cor:sample_bound}. We, however, note that the inclusion of derivative information still enhances the solution recovery.}
{\color{black}In practice, there is often error (or noise) in the evaluation of $u(\bm\Xi)$ and its derivatives, with the latter being more prone to errors. To model such inaccuracies, here we multiplying the realizations of $u(\bm\Xi)$ and its derivatives from (\ref{eqn:manufactured_u}) and (\ref{eqn:manufactured_upar}), respectively, by independent realizations of $(1+\epsilon_N)$, where $\epsilon_N$ is a zero mean Gaussian random variable with variance $10^{-5}$. In Fig.~\ref{fig:PCE_S50_noisy}, we consider the sparsity $|\mathcal{C}|=50$ case and compute the probability of successful solution recovery as a function of $\tilde{N}$. Similar as in the previous test cases, we use an RRMSE error of $0.01\%$ to identify a successful recovery. We observe that the inclusion of derivative information, while imprecise, still improves the performance of the standard $\ell_1$-minimization.}

\subsection{Case II: Two-dimensional elliptic PDE with random coefficient}
\label{subsec:PDE}
We next consider the two-dimensional (in space) elliptic PDE
\begin{equation}
\begin{aligned}
&-\nabla\cdot \left(a(\bm x,\bm{\Xi})\nabla u(\bm x,\bm{\Xi})\right)=1, &\quad &\bm x\in \mathcal{D}=[0,1]^2~,\\
&u(\bm x,\bm{\Xi})=0, &\quad &\bm x\in\partial\mathcal{D}~,
\end{aligned}
\label{eqn:PDE}
\end{equation}
where the diffusion coefficient $a(\bm x,\bm{\Xi})$ is modeled by the lognormal random field
\begin{equation}
a(\bm x, \bm\Xi) = \exp{\left[\bar{a}+\sigma_a\sum_{k=1}^d\sqrt{\lambda_k}\phi_k(\bm x)\Xi_k\right]}.
\label{eq:kl}
\end{equation}
Here, $d=30$, and $\Xi_k,~k=1,\dots,d$, are independent standard Gaussian random variables. In addition, $\bar{a}=0.1$, $\sigma_a=0.5$, and $\left\{\phi_k\right\}_{k=1}^d$ are the eigenfunctions corresponding to the $d$ largest eigenvalues $\left\{\lambda_k\right\}_{k=1}^d$ of the Gaussian covariance kernel
\begin{equation}
C_{aa}(\bm x_1, \bm x_2)=\exp\left[-\frac{(x_1-x_2)^2}{l_c^2}-\frac{(y_1-y_2)^2}{l_c^2}\right]~,
\end{equation}
with correlation length $l_c=1/16$. 

The QoI $u({\color{black}\bm\Xi})$ is chosen as the solution to \eqref{eqn:PDE} at location $\bm x=(0.5,0.5)$, i.e., $u({\color{black}\bm\Xi})=u\left((0.5,0.5),{\color{black}\bm\Xi}\right)$. The derivatives $\partial u/\partial\Xi_k$ are computed using the adjoint sensitivity method explained in detail in Section~\ref{adj:adj}. Both forward and adjoint solvers are implemented in the finite element method (FEM) project FEniCS \cite{FEniCS}.

\subsubsection{Adjoint sensitivity derivatives}
\label{adj:adj}

The adjoint sensitivity methods are commonly used in research areas such as sensitivity analysis~\cite{Sykes85, Cao03}, optimization~\cite{Giles00, Bryson75}, shape design~\cite{Laporte03, Jameson03}, etc., to compute the derivatives of the solutions of interest with respect to the underlying model parameters. In this work, we adopt the discrete adjoint sensitivity method to compute derivatives of the QoI at the $\bm\Xi$ samples.

{\color{black}In detail, for the interest of convenience, we consider the discrete formulation of the generic PDE in (\ref{eqn:PDE_operator}), at a fixed time, given by the residual equation}
\begin{equation}
\bm{\mathcal{R}}(\bm w,\bm\Xi)=\bm{0},
\label{eq:forward}
\end{equation}
where $\bm{w}\in\mathbb{R}^{M\times 1}$ contains the discrete values of solution over the spatial domain $\mathcal{D}$, and $M$ is the number of solution degrees-of-freedom. Recalling that $u(\bm\Xi)$ (here a scalar functional of the solution) denotes the QoI, we seek to compute the sensitivity derivatives $du/d\Xi_k$ from
\begin{equation}
\label{eqn:der_u}
\frac{du}{d\Xi_k} = \frac{\partial u}{\partial \Xi_k} + \frac{\partial u}{\partial \bm w}\frac{\partial \bm w}{\partial \Xi_k}.
\end{equation}
Taking the derivative of \eqref{eq:forward} with respect to $\Xi_k$, we have
\begin{equation}
\frac{\partial \bm{\mathcal{R}}}{\partial\Xi_k}+\frac{\partial \bm{\mathcal{R}}}{\partial\bm w}\frac{\partial\bm w}{\partial\Xi_k} = \bm 0,
\end{equation}
which results in $\partial\bm w/\partial\Xi_k =- \left(\partial \bm{\mathcal{R}}/\partial\bm w\right)^{-1}\partial \bm{\mathcal{R}}/\partial\Xi_k$. Plugging this in (\ref{eqn:der_u}) gives
\begin{equation}
\label{eqn:der_u1}
\frac{du}{d\Xi_k} = \frac{\partial u}{\partial \Xi_k} - \frac{\partial u}{\partial \bm w}\left(\frac{\partial \bm{\mathcal{R}}}{\partial\bm w}\right)^{-1}\frac{\partial \bm{\mathcal{R}}}{\partial\Xi_k},\nonumber
\end{equation}
which can be rewritten as
\begin{equation}
\label{eqn:der_u2}
\frac{du}{d\Xi_k} = \frac{\partial u}{\partial \Xi_k} + \bm{\lambda}^{T}\frac{\partial \bm{\mathcal{R}}}{\partial\Xi_k},
\end{equation}
where $\bm\lambda$ is the solution to the discrete adjoint equation 
\begin{equation}
\label{eqn:adjoint}
\left(\frac{\partial \bm{\mathcal{R}}}{\partial\bm w}\right)^T\bm\lambda = -\left(\frac{\partial u}{\partial \bm{w}}\right)^T.
\end{equation}

For the case of elliptic PDE (\ref{eqn:PDE}), $\partial \bm{\mathcal{R}}/\partial\bm w$ is the symmetric stiffness matrix of the FEM discretization and $\partial \bm{\mathcal{R}}/\partial\Xi_k$ in (\ref{eqn:der_u2}) can be computed semi-analytically from \eqref{eq:kl} and the FEM formulation of (\ref{eqn:PDE}). {\color{black}Here, we assume the inverse or a factorization of $\partial \bm{\mathcal{R}}/\partial\bm w$ is not stored when solving for $\bm w$, and that the total cost of obtaining $\partial \bm{\mathcal{R}}/\partial\Xi_k$ is smaller than that of computing $\bm w$. Therefore, the cost of solving for $\bm\lambda$ from (\ref{eqn:adjoint}) is roughly the same as solving for $\bm w$, which in turn suggests that $\nu=2$ in (\ref{eqn:tildeN}).}
\subsubsection{Results}
We approximate $u(\bm\Xi)$ in a Hermite PCE with total degree $p=3$, and seek to approximate the first $2500$ coefficients. We sort the elements of $\{\psi_j\}$ such that, for any given total order basis, the random variables $\Xi_k$ with smaller indices $k$ contribute first to the basis. 

\begin{figure}[H]
\centering
\subfloat[]{\label{fig:PDE_256_rms}\includegraphics[width=0.48\textwidth]{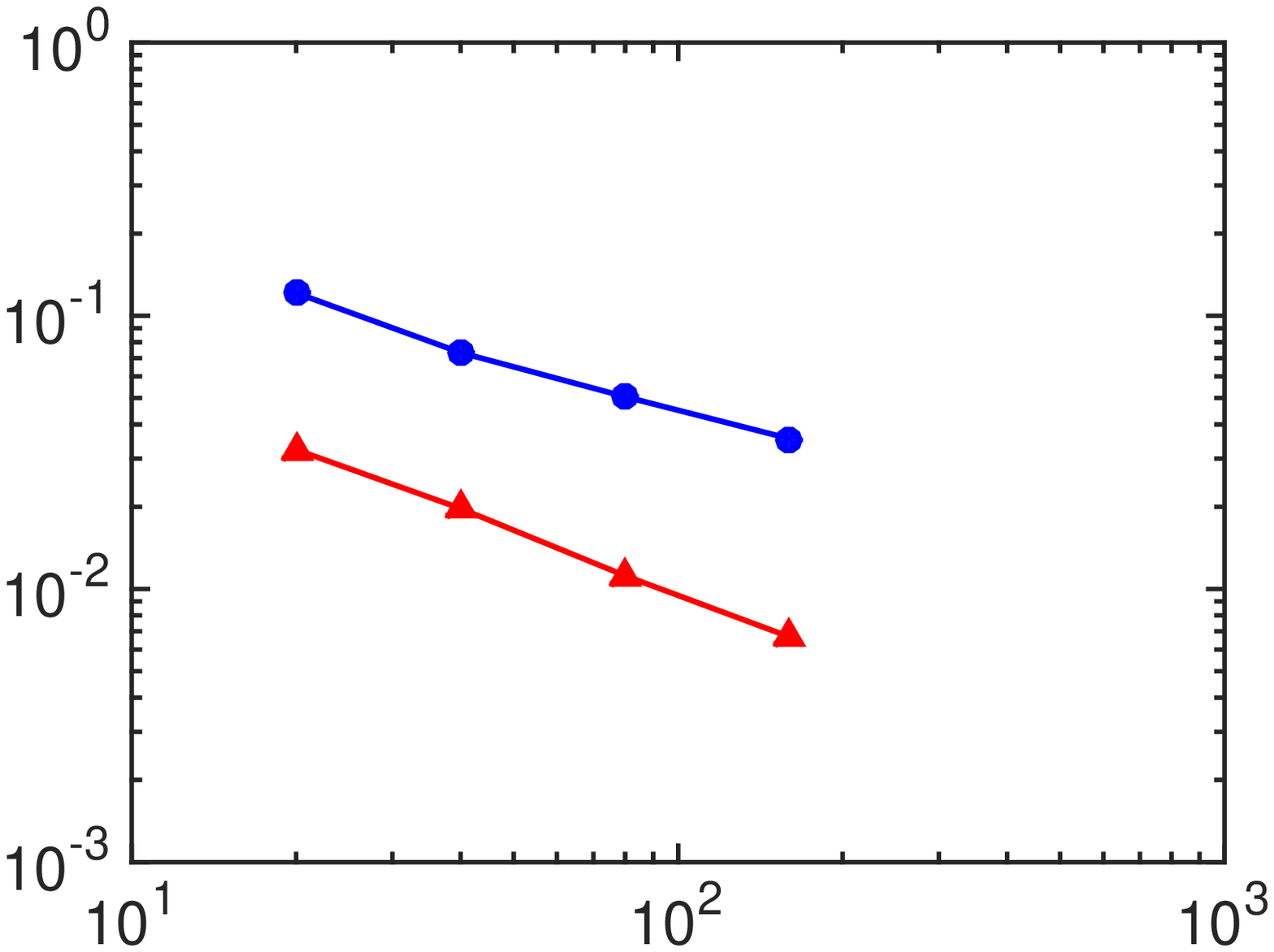}}
\put(-195,53){ \begin {sideways} \fontfamily{phv}\selectfont {\scriptsize Mean RRMSE} \end{sideways}}
\put(-96.5,0){{\scriptsize $\tilde{N}$}}
\hspace{.2cm}
\subfloat[]{\label{fig:PDE_256_std}\includegraphics[width=0.48\textwidth]{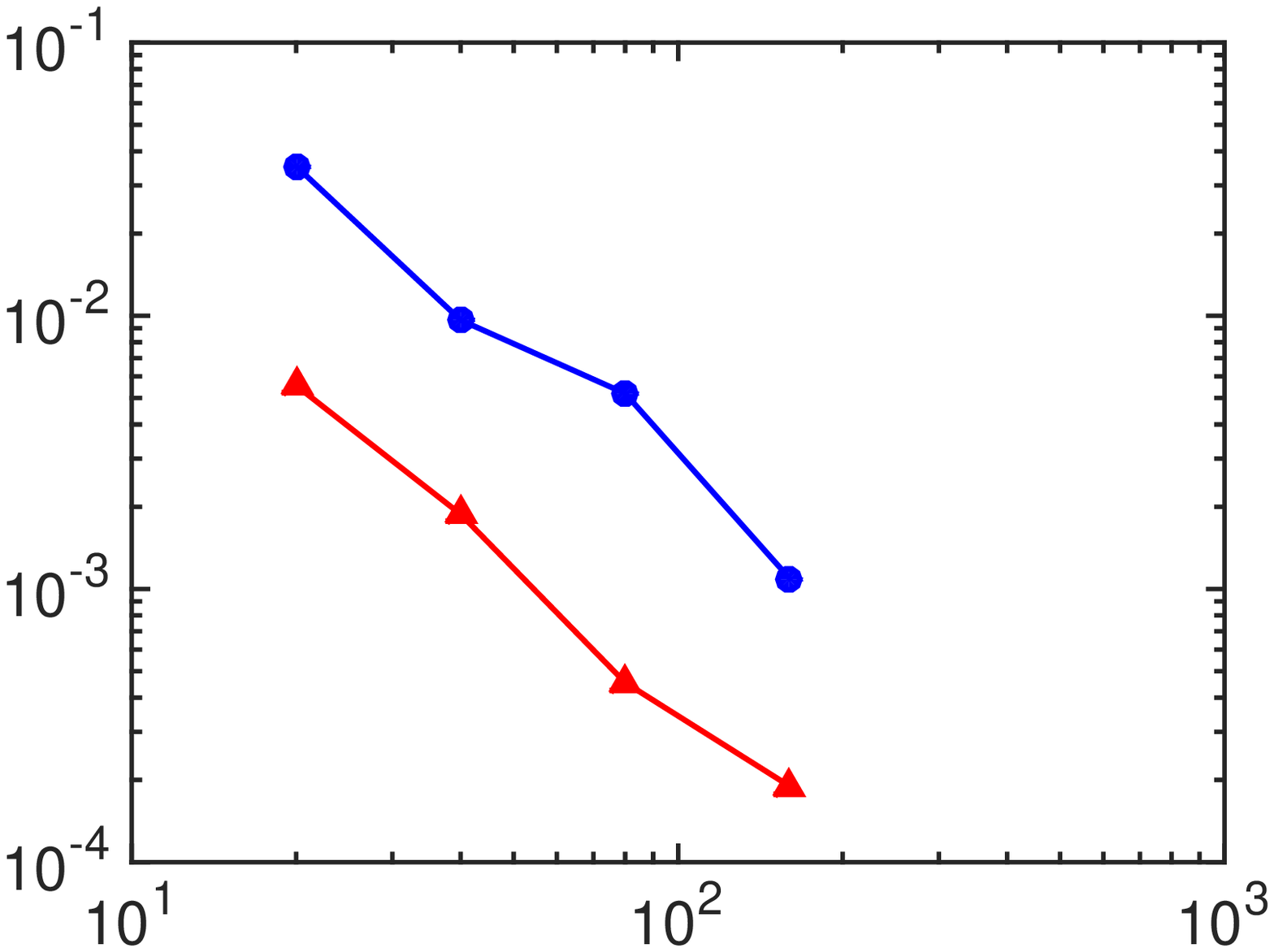}}
\put(-195,53){ \begin {sideways}\fontfamily{phv}\selectfont \scriptsize  Std of RRMSE \end{sideways}}
\put(-96.5,0){{\scriptsize $\tilde{N}$}}
\caption{Comparison of the statistics of the RRMSE in reconstructing $u\left((0.5,0.5),\bm\Xi\right)$, where the realizations of $u$ and its derivatives are computed on a uniform $256\times256$ FEM mesh. (a) Mean of RRMSE. (b) Standard deviation of RRMSE. (\usebox{\LegendeRDU}~100\% gradient-enhanced, \usebox{\LegendeBC}~standard)}
\label{fig:PDE_256}
\end{figure}

\begin{figure}[H]
\centering
\includegraphics[width=\textwidth]{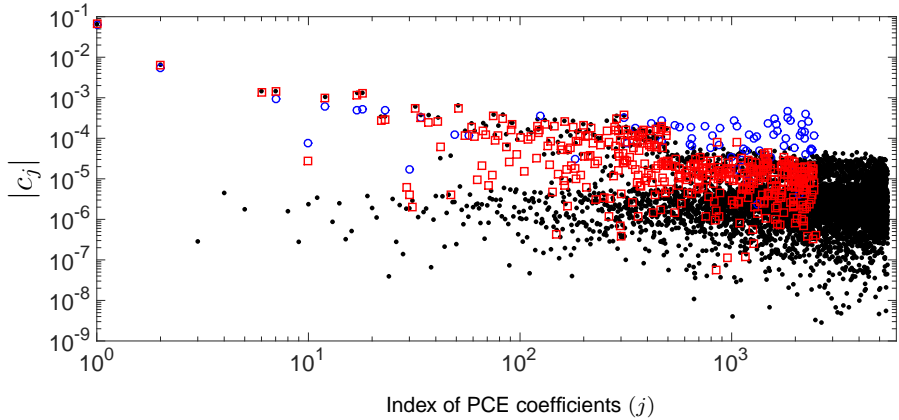}
\put(-377,70){ \begin {sideways}\fontfamily{phv}\selectfont   $\vert c_j\vert$ \end{sideways}}
\put(-230,-10){{\fontfamily{phv}\selectfont\scriptsize Index of PC{\color{black}E} coefficients $(j)$}}
\caption{Approximate PCE coefficients of $u\left((0.5,0.5),\bm\Xi\right)$ with $\tilde{N}=80$ vs. the reference coefficients obtained by least squares regression. ({$\bullet$} reference, {\color{blue} $\circ$} standard $\ell_1$-minimization, {\color{red} \scriptsize $\square$} gradient-enhanced $\ell_1$-minimization)}
\label{fig:PDE_coe}
\end{figure}

To solve for the coefficients, we first generate the realizations $\bm{u}$ and $\bm{u}_\partial$ using a $256\times256$ uniform, linear FEM mesh, which resolves both quantities with {\it low} numerical errors. %The mean normalized {\color{red}residuals in solving for} $u(\bm\xi^{(i)})$ and $\bm u_\partial(\bm{\xi}^{(i)})$ are $1.7\times 10^{-3}$ and $0.24$, respectively. 
In Fig.~\ref{fig:PDE_256}, we compare the mean and standard deviation of the RRMSE for solutions computed by the standard and gradient-enhanced $\ell_1$-minimization, using $100$ independent replications. The reference PCE coefficients are computed using least squares regression~\cite{Hampton15} with $10000$ solution realizations and yields a relative error of $0.36\%$ for $1000$ additional validation samples. From Fig.~\ref{fig:PDE_256}, we observe that higher accuracies are achieved by the gradient-enhanced $\ell_1$-minimization with the same number of samples $\tilde{N}$. In Figure~\ref{fig:PDE_coe}, we show the magnitude of the approximate PCE coefficients computed via standard and gradient-enhanced $\ell_1$-minimization with $\tilde{N}=80$ samples. More accurate coefficient estimates are obtained by the gradient-enhanced $\ell_1$-minimization.

\begin{figure}[H]
\centering
\subfloat[]{\label{fig:PDE_128_rms}\includegraphics[width=0.48\textwidth]{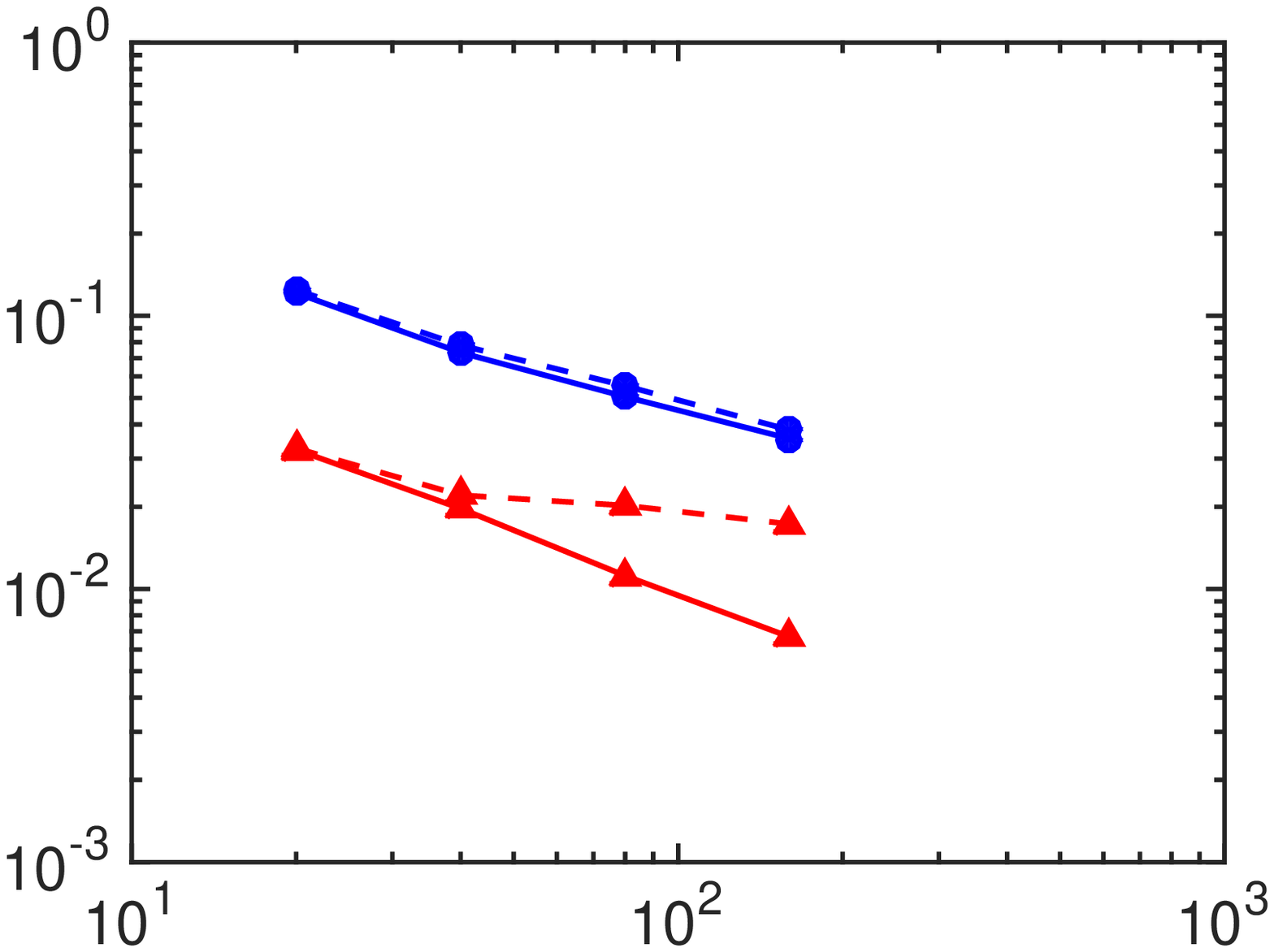}}
\put(-195,53){ \begin {sideways} \fontfamily{phv}\selectfont {\scriptsize Mean RRMSE} \end{sideways}}
\put(-96.5,0){{\scriptsize $\tilde{N}$}}
\hspace{.2cm}
\subfloat[]{\label{fig:PDE_128_std}\includegraphics[width=0.48\textwidth]{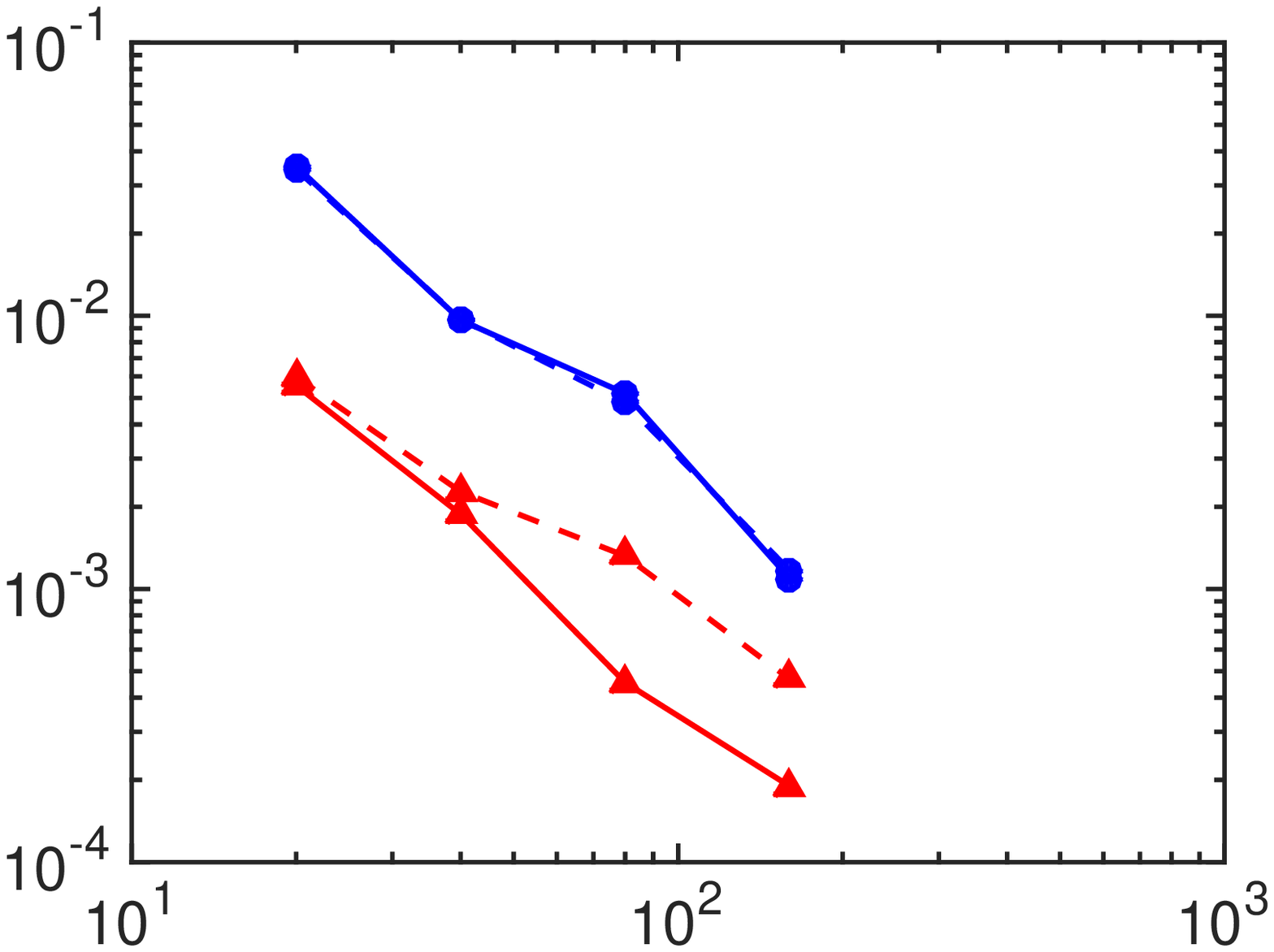}}
\put(-195,53){ \begin {sideways}\fontfamily{phv}\selectfont \scriptsize Std of RRMSE \end{sideways}}
\put(-96.5,0){{\scriptsize $\tilde{N}$}}
\caption{Comparison of the statistics of the RRMSE in reconstructing $u\left((0.5,0.5),\bm\Xi\right)$ via gradient-enhanced and standard $\ell_1$-minimization. (a) Mean of RRMSE. (b) Standard deviation of RRMSE. (\usebox{\LegendeRDU}~100\% gradient-enhanced, \usebox{\LegendeBC}~standard. Dashed and solid lines, respectively, correspond to the $16\times 16$ and $256\times 256$ mesh simulations.)}
\label{fig:PDE_128}
\end{figure}

To study how the accuracy of derivative information affects the accuracy of solution obtained by the gradient-enhanced $\ell_1$-minimization, we repeat this experiment on a coarser $16\times16$ mesh, where the derivative information is noticeably less accurate. %In this coarse mesh, the mean normalized residual {\color{red}in solving for} $u(\bm\xi^{(i)})$ is $0.39$, while $\bm u_\partial(\bm{\xi}^{(i)})$ is $824.40$. 
We observe from Fig.~\ref{fig:PDE_128} that, the accuracy improvement achieved by the gradient-enhanced $\ell_1$-minimization is not as considerable as in the case of $256\times 256$ mesh. This suggests that {\color{black}high} accuracy on the derivative samples $\bm{u}_{\partial}$ may be required for the gradient-enhanced $\ell_1$-minimization to be most effective. %Unlike the gradient-enhanced $\ell_1$-minimization, the standard $\ell_1$-minimization exhibits less sensitivity to the errors in the evaluation of $\bm u$.} 

\subsection{Case III: Plane Poiseuille flow with random boundaries}
\label{subsec:Poiseuille}
\begin{figure}[H]
\vspace{.5cm}
\includegraphics[scale=1]{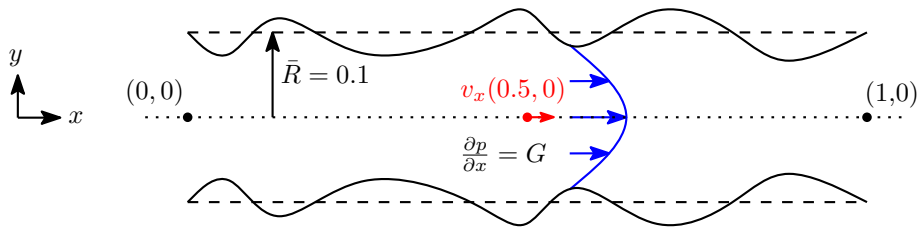}
\caption{Schematic figure of plane Poiseuille flow with random boundaries.}
\label{fig:wavychannel}
\end{figure}

We consider a 2-D plane Poiseuille flow with random boundaries as depicted in Fig.~\ref{fig:wavychannel}. 

\vspace{.3cm}

The average width of the channel is $2\bar{R}=0.2$, and the width of the channel is model as $2R(x)=2\bar{R}+2r(x,\bm\Xi)$, where $2r(x,\bm\Xi)$ describes the random fluctuation of the channel width around $2\bar{R}$. We use $d=20$ independent standard Gaussian random variables $\Xi_k,k=1,\dots,d$, to represent the uncertainty in $R$, and let
\begin{equation}
r(x,\bm\Xi) = \exp\left[\bar{r}+\sigma_r\sum_{k=1}^d\sqrt{\lambda_k}\phi_k(x)\Xi_k\right].
\end{equation}
Here, $\bar{r}=-4$, $\sigma_r=0.5$, and $\left\{\lambda_k\right\}_{k=1}^d$ and $\left\{\phi_k(x)\right\}_{k=1}^d$ are the $d$ largest eigenvalues and corresponding eigenfunctions of the exponential covariance kernel
\begin{equation}
C_{rr}(x_1,x_2) = \exp\left(-\frac{|x_1-x_2|}{l_c}\right)~,
\end{equation}
where the correlation length $l_c=1/21$.

We seek to investigate the steady state velocity field of the flow, which is governed by the incompressible steady state Navier-Stokes equations
%
%\begin{equation}
\begin{align}
&(\bm v\cdot \nabla)\bm v - \frac{1}{Re}\nabla^2\bm v=-\nabla p, &\quad &\bm x\in\mathcal{D}(\bm{\Xi}),\label{eq:NavierStokes}\\
&\nabla\cdot \bm v = 0, &\quad & \bm x\in\mathcal{D}(\bm{\Xi}),\nonumber\\
&\frac{\partial p}{\partial x} = G,&\quad &\bm x\in\mathcal{D}(\bm{\Xi}),\nonumber\\
& \bm v = 0, &\quad & y=\pm R,\label{eq:NavierStokes_end}\nonumber
\end{align}
%\end{equation}
%
where the Reynolds number $Re=60$, and $p$ denotes pressure. Notice that in \eqref{eq:NavierStokes}, we assume that all physical quantities are non-dimensional. The flow is driven by a pressure gradient $G=-0.1$.

\begin{figure}[H]
\centering
\includegraphics[width=0.75\textwidth]{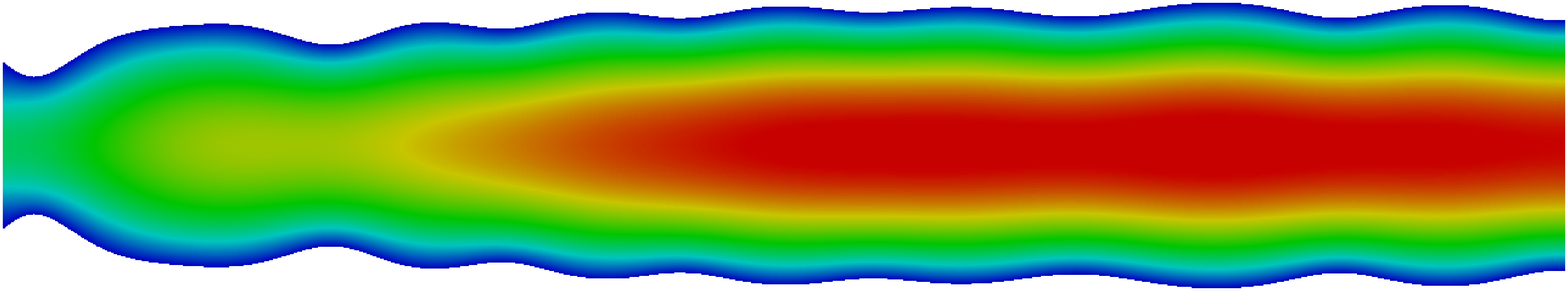}\includegraphics[scale=0.25]{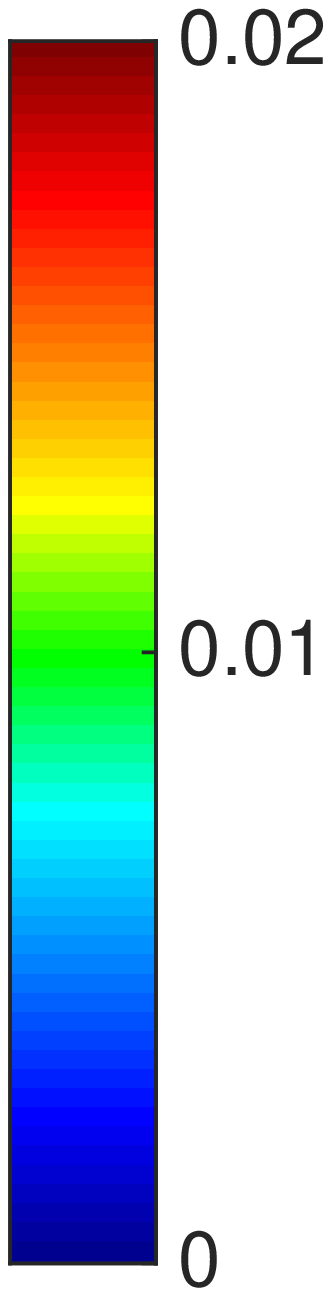}
\includegraphics[width=0.75\textwidth]{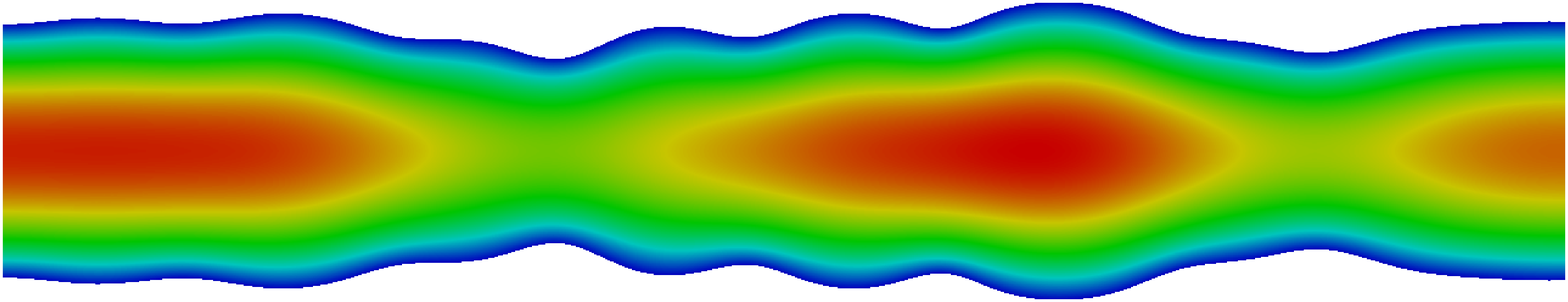}\includegraphics[scale=0.25]{flow/colormap.eps}
\caption{Velocity magnitude contours with two independent realizations of the random inputs.}
\label{fig:vel_field}
\end{figure}

The QoI is $v_x(0.5,0)$, the horizontal velocity at $(0.5,0)$, and we approximate it in a  Hermite PCE of total degree $p=3$.

\subsubsection{Adjoint sensitivity derivatives}

To compute the derivative information, we again adopt the adjoint sensitivity method, in which we approximate ${\partial \bm{\mathcal R}}/{\partial \Xi_k},~k=1,\dots,d$, in \eqref{eqn:der_u2} via finite difference quotient
\begin{equation}
\frac{\partial \bm{\mathcal R}}{\partial\Xi_k}\approx\frac{\bm{\mathcal R}(\bm w,\bm\Xi+
\Delta\bm\Xi^k)-\bm{\mathcal R}(\bm w,\bm\Xi)}{\varepsilon}.
\label{eq:dLdxi}
\end{equation}
Here, the $m$th entry of the vector $\Delta\bm\Xi^k$ is defined as
\begin{equation}
\Delta\Xi^k_m=
\left\{
\begin{aligned}
&\varepsilon, & \quad & m=k,\\
&0, & \quad & m\ne k.\\
\end{aligned}
\right.
\end{equation}
To solve the non-linear problem (\ref{eq:NavierStokes}), we employ a standard Newton solver, in which $\partial\bm{\mathcal R}/{\partial \bm w}$ from \eqref{eqn:adjoint} is the Jacobian matrix. In \eqref{eq:dLdxi}, the discrete representation of the velocity field, $\bm w$, is kept unchanged; hence, \eqref{eq:NavierStokes} does not need to be solved again. The perturbed residual $\bm{\mathcal R}(\bm w,\bm\Xi+\Delta\bm\Xi^k)$ is computed by deforming the mesh to conform to the geometry corresponding to $\bm\Xi+\Delta\bm\Xi^k$, without recomputing $\bm w$. Compared to solving the adjoint equation \eqref{eqn:adjoint}, the cost of calculating \eqref{eq:dLdxi} is negligible. Additionally, in our deterministic solver, computing $\bm w$ required on average 3 Newton steps. Therefore, the extra cost of computing derivative information is roughly equivalent to $1/3$ of the cost of computing $\bm w$, which in turn suggests setting $\nu=4/3$ in (\ref{eqn:tildeN}).

\subsubsection{Results}
We consider 0\%, 20\% and 100\% gradient-enhanced $\ell_1$-minimization, with  $\tilde{N}\in\left\{20,40,80,160\right\}$. 
\begin{figure}[H]
\centering
\subfloat[]{\label{fig:flow_rms}\includegraphics[width=0.48\textwidth]{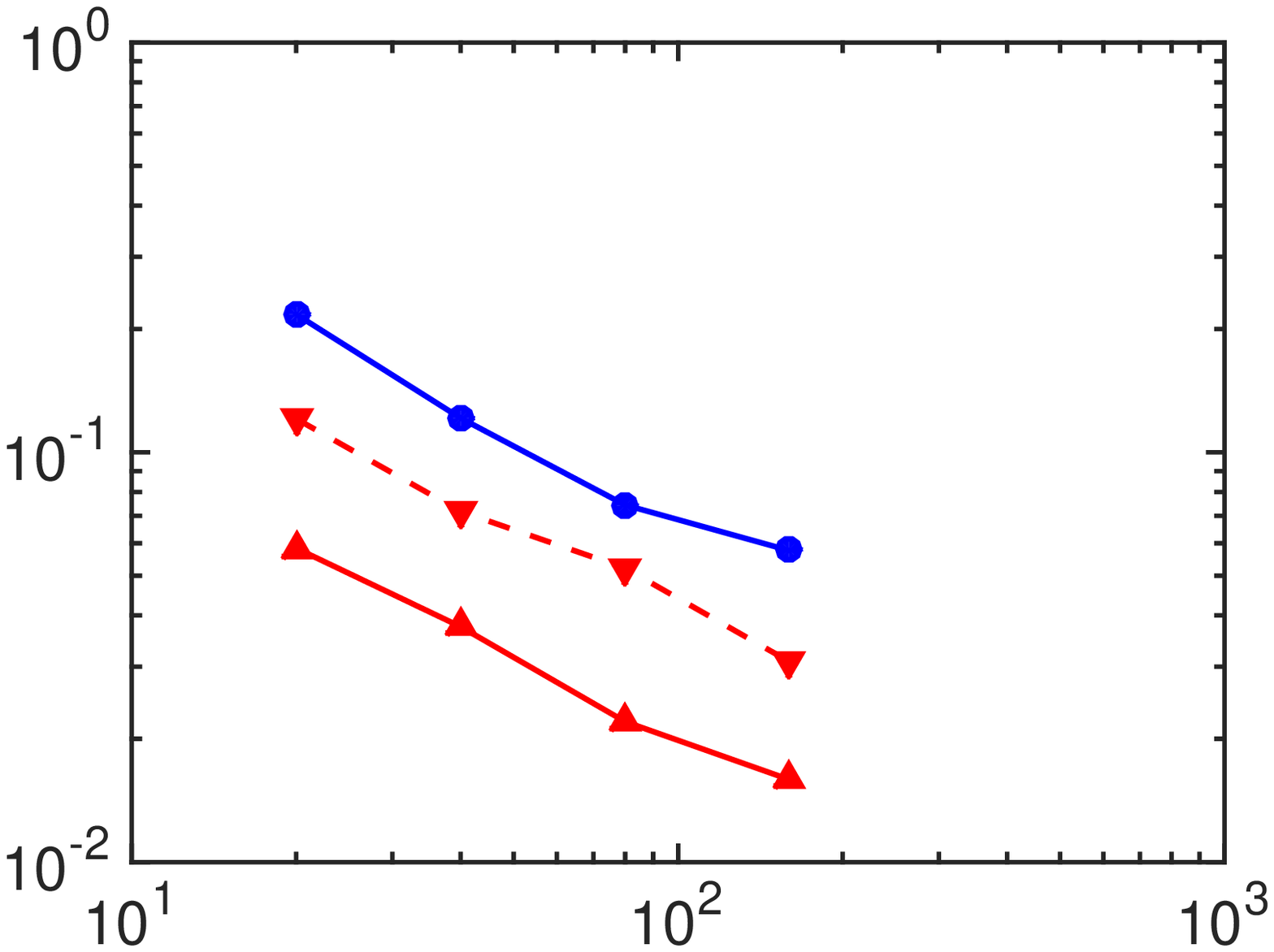}}
\put(-195,53){ \begin {sideways} \fontfamily{phv}\selectfont {\scriptsize Mean RRMSE} \end{sideways}}
\put(-96.5,0){{\scriptsize $\tilde{N}$}}
\hspace{.2cm}
\subfloat[]{\label{fig:flow_std}\includegraphics[width=0.48\textwidth]{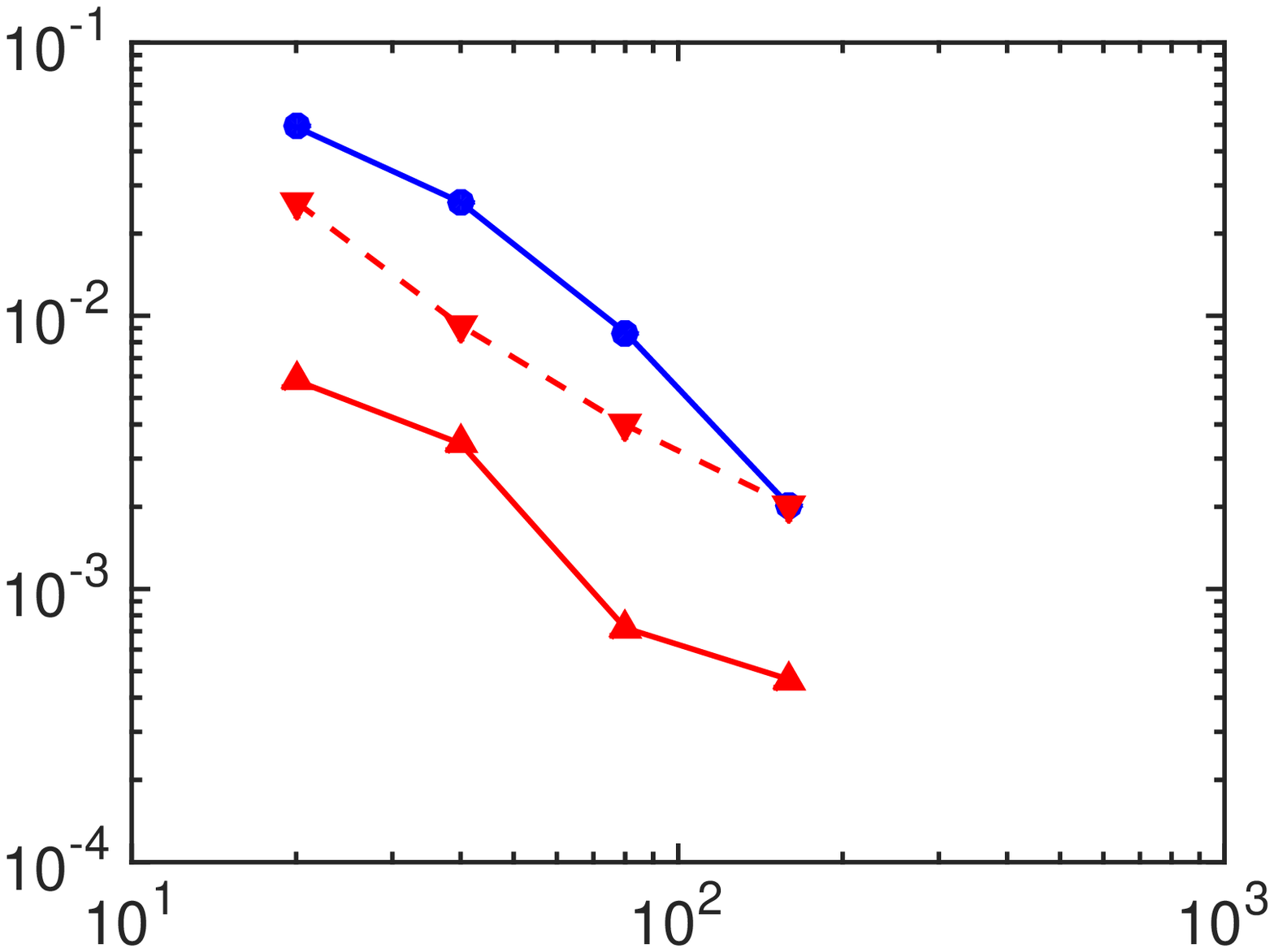}}
\put(-195,53){ \begin {sideways}\fontfamily{phv}\selectfont \scriptsize Std of RRMSE \end{sideways}}
\put(-96.5,0){{\scriptsize $\tilde{N}$}}
\caption{Comparison of the statistics of the RRMSE in reconstructing $v_x\left((0.5,0),\bm\Xi\right)$ via standard and gradient-enhanced $\ell_1$-minimization, with $100$ independent replications. (a) Mean of RRMSE. (b) Standard deviation of RRMSE. (\usebox{\LegendeRDU}~$100\%$ gradient-enhanced, \usebox{\LegendeRDDD}~$20\%$ gradient-enhanced, \usebox{\LegendeBC}~standard)  }
%\caption{(a) Mean RRMSE. (b) Standard deviation of RRMSE. (\usebox{\LegendeRDU}~$100\%$ gradient-enhanced, \usebox{\LegendeRDDD}~$20\%$ gradient-enhanced, \usebox{\LegendeBC}~standard).}
\label{fig:flow_all}
\end{figure}
Fig.~\ref{fig:flow_all} displays the comparisons of the mean and standard deviation of the RRMSE, with $100$ independent replications of computed PCE coefficients, showing that gradient-enhanced $\ell_1$-minimization again leads to cost-effective accuracy improvement.

\section{Proofs}
\label{sec:proofs}
\subsection{Theorem~\ref{thm:mu_beta}}
We prove our results here in the case of $\beta_{\mathcal{Q}}$ defined as in (\ref{eqn:matrix_coh_def}), which due to the non-independent rows is a slight generalization of the results using (\ref{eqn:std_coh_def}). Adjusting the proofs present here to account for the latter case requires only the substitution of $\mu_{\mathcal{Q}}$ for $\beta_{\mathcal{Q}}$, showing this Theorem.
\subsection{Lemma~\ref{lem:hermite_normalize}.}
We begin by providing a brief proof for Lemma~\ref{lem:hermite_normalize}, which follows directly from the explicit form for the Hermite derivative as in (5.5.10) of~\cite{Szego}.

\proof Note that linearity of expectation allows us to take the expectation inside the sum, so that we may work on each term independently. For $1$-dimensional orthonormal Hermite polynomials, where $i$ represents the order of the polynomial (5.5.10) of~\cite{Szego} shows that
\begin{align}
\label{eqn:herm_prime}
\frac{\partial\psi_i}{\partial\Xi} (\Xi) = \sqrt{i}\psi_{i-1}(\Xi).
\end{align}
As the tensor product of orthonormal polynomials is orthonormal, (\ref{eqn:herm_norm}) for $\bm{i}=\bm{j}$ follows from the derivative being exactly in the direction of a Hermite polynomial. For $\bm{i}\ne\bm{j}$, (\ref{eqn:herm_norm}) follows because each term in the sum is the expectation of a product of orthogonal polynomials that differ in at least one coordinate, and hence the integral is equal to zero. $\blacksquare$
\subsection{Theorem~\ref{thm:1}}
To prove Theorem~\ref{thm:1}, we appeal to a matrix variant of the Chernoff bound~\cite{Tropp12a}, which is similar to approaches taken in~\cite{CDL13,CandesPlan,Baraniuk08,Hampton15}.
\proof
Recall that $\mathcal{Q}$ represents a truncation for the domain of $\bm{\Xi}$, and may be given by (\ref{eqn:trunc_def}). We have that
\begin{align}
\bm{I} &=\mathbb{E}\left(\bm{X}^T\bm{X}|\bm\xi\in\mathcal{Q}\right)\mathbb{P}(\mathcal{Q})+\mathbb{E}\left(\bm{X}^T\bm{X}|\bm\xi\in\mathcal{Q}^c\right)\mathbb{P}(\mathcal{Q}^c).
\end{align}
A brief calculation gives that
\begin{align}
\label{eqn:epsdef}
\epsilon_{\mathcal{Q}}&:=\left\|\mathbb{E}\left(\bm{X}^T\bm{X}|\bm\xi\in\mathcal{Q}\right)-\bm{I}\right\|_2\\
&\le\frac{\mathbb{P}(\mathcal{Q}^c)}{\mathbb{P}(\mathcal{Q})}\left(\left\|\mathbb{E}\left(\bm{X}^T\bm{X}\bigg|\bm\xi\in\mathcal{Q}^c\right)\right\|_2 + 1\right),
\end{align}
bounds the bias introduced from not accepting samples within $\mathcal{Q}^c$. We note that with the truncation in (\ref{eqn:trunc_def}), and using this bound, $\epsilon_{\mathcal{Q}}\le 0.1/\sqrt{P}$~\cite{Hampton14}, for all $N$ considered here, and in most similar problems. Specifically, an analytic issue does not arise until consideration of exponential levels of oversampling. This analytic issue concerns with the truncated rare events being reliably observed due to the very large sample pool.

Restating (\ref{eqn:epsdef}), let $\lambda_{\min}(\cdot)$ and $\lambda_{\max}(\cdot)$ correspond to the smallest and largest eigenvalues of the argument matrix, respectively. Then for any arbitrary set of columns, denoted $\mathcal{S}$,
\begin{align} 
1-\epsilon_{\mathcal{Q}}&\le\lambda_{\min}\left(\mathbb{E}\left(\bm{X}^T(:,\mathcal{S})\bm{X}(:,\mathcal{S})|{\color{black}\bm{\xi}}\in\mathcal{Q}\right)\right)\\
&\le\lambda_{\max}\left(\mathbb{E}\left(\bm{X}^T(:,\mathcal{S})\bm{X}(:,\mathcal{S})|{\color{black}\bm{\xi}}\in\mathcal{Q}\right)\right)\le1+\epsilon_{\mathcal{Q}}.
\end{align}
From (\ref{eqn:matrix_coh_def}) we have that if each sample {\color{black}$\bm{\xi}^{(k)}\in\mathcal{Q}$}, then for all $k$,
\begin{align}
\label{eqn:matrix_fro_bound}
\|\bm{X}_k(:,\mathcal{S})\|^2_2&\le s\beta_{\mathcal{Q}},
\end{align}
holds uniformly for all choices of $\mathcal{S}$ such that $|\mathcal{S}|<s$. This provides an upper bound on the singular values of our independent self-adjoint matrices, $\bm{X}_k^T(:,\mathcal{S})\bm{X}_k(:,\mathcal{S})$, uniformly over all choices of $\mathcal{S}$ consisting of at most $s$ elements. We define
\begin{align}
\bm{M}_{\mathcal{S}}:=\frac{1}{N}\mathop{\sum}\limits_{k=1}^N\bm{X}_k^T(:,\mathcal{S})\bm{X}_k(:,\mathcal{S}).
\end{align}
An application of the Chernoff bound as in Theorem 1.1 of~\cite{Tropp12a} and Theorem 1 of~\cite{CDL13} gives that for $\delta \in[0,1]$ and $|\mathcal{S}|\le s$,
\begin{align}
 \mathbb{P}\Bigg(\lambda_{\min}\left(\bm{M}_{\mathcal{S}}\right)\le (1-\delta)(1-\epsilon_{\mathcal{Q}})\Bigg| {\color{black}\bm{\xi}^{(k)}}\in\mathcal{Q}\ \forall k\Bigg)&\le s\left(\frac{e^{-\delta}}{(1-\delta)^{1-\delta}}\right)^{\frac{N(1-\epsilon_{\mathcal{Q}})}{s\beta_{\mathcal{Q}}}};\\ 
\mathbb{P}\Bigg(\lambda_{\max}\left(\bm{M}_{\mathcal{S}}\right)\ge (1+\delta)(1+\epsilon_{\mathcal{Q}})\Bigg| {\color{black}\bm{\xi}^{(k)}}\in\mathcal{Q}\ \forall k\Bigg)&\le  s\left(\frac{e^{\delta}}{(1+\delta)^{1+\delta}}\right)^{\frac{N(1+\epsilon_{\mathcal{Q}})}{s\beta_{\mathcal{Q}}}}. 
\end{align}
Note that
\begin{align}
 (1-\delta)(1-\epsilon_{\mathcal{Q}})\ge 1-t &\implies \delta \le \frac{t-\epsilon_{\mathcal{Q}}}{1-\epsilon_{\mathcal{Q}}};\\
 (1+\delta)(1+\epsilon_{\mathcal{Q}})\le 1+t &\implies \delta \le \frac{t-\epsilon_{\mathcal{Q}}}{1+\epsilon_{\mathcal{Q}}}. 
\end{align}
and so we have  a critical $\delta$, given by
\begin{align}
\delta_{t}:=(t-\epsilon_{\mathcal{Q}})/(1+\epsilon_{\mathcal{Q}}),
\end{align}
is such that for all $\delta<\delta_{t}$ the matrix $\bm{M}_{\mathcal{S}}$ is guaranteed to satisfy $\|\bm{M}_{\mathcal{S}}-\bm{I}\|_2\le t$. Note that for $0\le\delta<1$,
\begin{align}
 \frac{e^{-\delta}}{(1-\delta)^{1-\delta}}\ge\frac{e^{\delta}}{(1+\delta)^{1+\delta}},
\end{align}
and so we may bound the sum of the probabilities by
\begin{align}
  \mathbb{P}\Bigg(\|\bm{M}_{\mathcal{S}}-\bm{I}\|\le t\Bigg| {\color{black}\bm{\xi}^{(k)}}\in\mathcal{Q}\ \forall k\Bigg)\le 2s\left(\frac{e^{-\delta_t}}{(1-\delta_t)^{1-\delta_t}}\right)^{\frac{N(1-\epsilon_{\mathcal{Q}})}{s\beta_{\mathcal{Q}}}}.
\end{align}
We now bound this probability. We note that if $\epsilon_{\mathcal{Q}}\le t$ then
\begin{align}0<
\delta_t\le\frac{t-\epsilon_{\mathcal{Q}}}{1+\epsilon_{\mathcal{Q}}}\le t-\epsilon_{\mathcal{Q}}.
\end{align}
To create a bound without explicit dependence on $\epsilon_{\mathcal{Q}}$, we note that for 
\begin{align*}
c_t&:=t-\epsilon_{\mathcal{Q}} + (t+\epsilon_{\mathcal{Q}})\log(t+\epsilon_{\mathcal{Q}});\\
 \frac{e^{-\delta_t}}{(1-\delta_t)^{1-\delta_t}}&\le\exp(-c_t).
\end{align*}
Thus for $t\in (0,1)$,
\begin{align}
\mathbb{P}\Bigg(\|\bm{M}_{\mathcal{S}}-\bm{I}\|\le t\Bigg| {\color{black}\bm{\xi}^{(k)}}\in\mathcal{Q}\ \forall k\Bigg)&\le 2s\exp\left(-\frac{c_t(1-\epsilon_{\mathcal{Q}})N}{s\beta_{\mathcal{Q}}}\right);\\
&\le 2s\exp\left(-C_{\mathcal{Q}}\frac{tN}{s\beta_{\mathcal{Q}}}\right),
\end{align}
where $C_{\mathcal{Q}}$ is a reasonably large positive constant for most truncations.
Via a union bound over the ${P \choose s}$ possibilities of subsets of $P$ with cardinality $s$, it follows that
\begin{align}
\mathbb{P}\left(\mathop{\sup}\limits_{|\mathcal{S}|\le s}\lambda_{\max}(\bm{M}_{\mathcal{S}}-\bm{I})\ge t\right) \le 2s{P \choose s}\exp\left(-C_{\mathcal{Q}}\frac{Nt}{s\beta_{\mathcal{Q}}}\right).
\end{align}
Recalling that 
\begin{align*}
\mathop{\sup}\limits_{|\mathcal{S}|\le s}\lambda_{\max}(\bm{M}_{\mathcal{S}}-\bm{I})=\delta_s,
\end{align*}
gives
\begin{align}
\mathbb{P}(\delta_s\ge t) \le \exp\left(-C_{\mathcal{Q}}\frac{Nt}{s\beta_{\mathcal{Q}}}+\log\left(2s{P\choose s}\right)\right).
\end{align}
We assume that having any sample ${\color{black}\bm{\xi}^{(k)}}\in\mathcal{Q}^{c}$ leads to an arbitrarily large $\lambda_{\max}$, hence yielding the bound,
\begin{align}
\mathbb{P}(\delta_s\ge t) \le 1 - \mathbb{P}(\mathcal{Q})^N +\exp\left(-C_{\mathcal{Q}}\frac{Nt}{s\beta_{\mathcal{Q}}}+\log\left(2s{P\choose s}\right)\right).
\end{align}
Using the relation, $\mathbb{P}(\delta_s< t)=1-\mathbb{P}(\delta_s \ge t)$, gives that
\begin{align}
\mathbb{P}(\delta_s< t) \ge \mathbb{P}(\mathcal{Q})^N-\exp\left(-C_{\mathcal{Q}}\frac{Nt}{s\beta_{\mathcal{Q}}}+\log\left(2s{P\choose s}\right)\right).
\end{align}
Using the relation that
\begin{align*}
{P\choose s}\le \left(\frac{eP}{s}\right)^s,
\end{align*}
it follows that
\begin{align}
2s{P\choose s} &\le 2se^s\left(\frac{P}{s}\right)^{s};\\
\log\left(2s{P\choose s}\right)&\le \log(2s) + s + s\log(P/s),
\end{align}
which completes the proof. Remark~\ref{rem:ell2} follows from taking $s=P$ and using that ${P\choose P} = 1$. $\blacksquare$

We note that Corollary~\ref{cor:sample_bound} follows from Theorem~\ref{thm:1}, by substituting $p_{\star}$ for $\mathbb{P}(\delta_s< t)$; substituting $\delta_{\star}$ for $t$; and performing some algebraic manipulation.
\subsection{Theorem~\ref{thm:2}}
The proof of Theorem~\ref{thm:2} relies on the properties of the measurement matrices $\tilde{\bm{\Psi}}$ and $\bm{\Psi}$ themselves. In an intuitive sense, the results follow from the two matrices having similar properties, but the gradient-enhanced matrix having more rows, yielding better conditioned Gramians.
\proof
We begin by showing {\bf R1} for the one-dimensional case. Note that for arbitrary ${\color{black}\xi}$, and $i$,
\begin{align*}
\frac{|\psi_i({\color{black}\xi})|^2 + i|\psi_{i-1}({\color{black}\xi})|^2}{1+i} &\le \max\left\{|\psi_i({\color{black}\xi})|^2,|\psi_{i-1}({\color{black}\xi})|^2\right\}
\end{align*}
and that equality can only hold if $\psi_i({\color{black}\xi})=\psi_{i-1}({\color{black}\xi})$ for some $i$, which is an event that occurs with probability zero. As this inequality holds for all $i$ and ${\color{black}\xi}$, being strict for almost all ${\color{black}\xi}$, {\bf R1} follows for the one-dimensional case. The $d$-dimensional analogue follows as the $d$-dimensional polynomials are tensor-products of the one-dimensional polynomials. 

To show {\bf R2}, note that up to an invertible pre-multiplication, $\bm{\Psi}$ is a submatrix of $\tilde{\bm{\Psi}}$, and thus $\mathcal{N}(\bm{\Psi})\subset \mathcal{N}(\tilde{\bm{\Psi}})$. Additionally, we notice that $\bm{\Psi}$, and $\tilde{\bm{\Psi}}$ are almost surely full rank matrices.

We next show {\bf R3} in the case of $1$-dimensional Hermite polynomials. Here subscripts of matrices refer to the column corresponding to that polynomial order. We have by the normalization (\ref{eqn:weight_def}) and (\ref{eqn:herm_prime}) that
\begin{align}
\label{eqn:mutual_coherence_decomposition}
(\tilde{\bm{\Psi}}_i,\tilde{\bm{\Psi}}_j)&=\frac{(\bm{\Psi}_i,\bm{\Psi}_j) + \sqrt{ij}(\bm{\Psi}_{i-1},\bm{\Psi}_{j-1})}{\sqrt{(1+i)(1+j)}}.
\end{align}
It follows that,
\begin{align*}
|(\tilde{\bm{\Psi}}_i,\tilde{\bm{\Psi}}_j)|&\le\frac{|(\bm{\Psi}_i,\bm{\Psi}_j)| + \sqrt{ij}|(\bm{\Psi}_{i-1},\bm{\Psi}_{j-1})|}{\sqrt{(1+i)(1+j)}}.
\end{align*}
Applying a supremum,
\begin{align*}
\mathop{\sup}\limits_{i\ne j}|(\tilde{\bm{\Psi}}_i,\tilde{\bm{\Psi}}_j)|&\le\mathop{\sup}\limits_{i\ne j}\frac{|(\bm{\Psi}_i,\bm{\Psi}_j)| + \sqrt{ij}|(\bm{\Psi}_{i-1},\bm{\Psi}_{j-1})|}{\sqrt{(1+i)(1+j)}},\\
&\le\mathop{\sup}\limits_{i\ne j}|(\bm{\Psi}_{i},\bm{\Psi}_{j})|\frac{1 + \sqrt{ij}}{\sqrt{(1+i)(1+j)}},\\
\end{align*}
where we have used the inequality
\begin{align}
\label{eqn:mutual_coherence_inequality}
 \frac{1 + \sqrt{(i-1)(j-1)}}{\sqrt{(1+(i-1))(1+(j-1))}}&\le \frac{1 + \sqrt{ij}}{\sqrt{(1+i)(1+j)}}.
\end{align}
This shows the result for the one-dimensional case. The $d$-dimensional case leads to a decomposition in (\ref{eqn:mutual_coherence_decomposition}) with inner products of lower order along each dimension, and the inequality in (\ref{eqn:mutual_coherence_inequality}), is replaced by $d$ similar inequalities.$\blacksquare$
\section{Conclusion}
\label{sec:conclusion}
Within the context of compressive sampling of sparse polynomial chaos expansions, we investigated $\ell_1$-minimization when derivative information of a quantity of interest (QoI) is present. We provided analysis on gradient-enhanced $\ell_1$-minimization for Hermite polynomial chaos, in which we showed that, for a given normalization, including derivative information will not reduce the stability of the $\ell_1$-minimization problem. Further, we identified a coherence parameter that we used to bound the associated Restricted Isometry Constant, a useful and well-studied measure of the stability for solutions recovered by $\ell_1$-minimization as in the context used here.

Furthermore, we observed improved solution accuracy from {\color{black}gradient-enhanced $\ell_1$-minimization} in three numerical examples: Manufactured polynomials; an elliptic equation; and a {\color{black}plane Poiseuille flow with random boundaries}. Consistently, {\color{black}gradient-enhanced $\ell_1$-minimization} was seen to improve the quality of solution recovery {\color{black}at the same computational cost, or equivalently achieve the same solution quality at a reduced computational cost}. {\color{black}As the QoI derivatives are often more sensitive to discretization errors than the QoI itself, so too is the accuracy of the solution obtained by the gradient-enhanced $\ell_1$-minimization. This was empirically observed in the second numerical example considered, thereby suggesting high accuracy requirements on derivative calculations for the gradient-enhanced $\ell_1$-minimization to be most effective.} 

\section*{Acknowledgements}
This material is based upon work supported by the U.S. Department of Energy Office of Science, Office of Advanced Scientific Computing Research, under Award Number DE-SC0006402, and NSF grants DMS-1228359 and CMMI-1454601.

This work utilized the Janus supercomputer, which is supported by the National Science Foundation (award number CNS-0821794) and the University of Colorado Boulder. The Janus supercomputer is a joint effort of the University of Colorado Boulder, the University of Colorado Denver and the National Center for Atmospheric Research.

\section*{References}

\bibliographystyle{unsrt}	% or "siam", or "alpha", etc.
%\nocite{*}		% list all refs in database, cited or not
\bibliography{refs}	

\end{document}